\documentclass[pra,superscriptaddress,twocolumn,floatfig,showpacs,floatfix]{revtex4-1}
\usepackage{graphicx,color}
\usepackage{amsmath,amssymb,bm}
\usepackage{braket}		

\usepackage{mathtools}

\usepackage[plainpages=false,pdfpagelabels,colorlinks=true,linkcolor=red,urlcolor=blue,citecolor=blue,pdftitle={Eigenstate thermalization hypothesis through the lens of autocorrelation functions},pdfauthor={},pdfdisplaydoctitle=true,pdfduplex=DuplexFlipLongEdge]{hyperref}

\definecolor{darkred}{rgb}{0.90,0,0}
\definecolor{darkgreen}{rgb}{0,0.60,.2}
\definecolor{darkblue}{rgb}{0,0,1}
\definecolor{grey}{cmyk}{0,0,0,0.25}
\definecolor{orange}{cmyk}{0,0.6,1,0}

\begin{document}
\title{
Eigenstate thermalization hypothesis through the lens of autocorrelation functions
}

\author{Christoph Sch\"{o}nle}
\author{David Jansen}
\author{Fabian Heidrich-Meisner}
\affiliation{Institut f\"{u}r Theoretische Physik, Georg-August-Universit\"at G\"ottingen, D-37077 G\"ottingen, Germany}
\author{Lev Vidmar}
\affiliation{Department of Theoretical Physics, J. Stefan Institute, SI-1000 Ljubljana, Slovenia}
\affiliation{Department of Physics, Faculty of Mathematics and Physics, University of Ljubljana, SI-1000 Ljubljana, Slovenia}

\begin{abstract}
Matrix elements of observables in eigenstates of generic Hamiltonians are described by the Srednicki ansatz within the eigenstate thermalization hypothesis (ETH).
We study a quantum chaotic spin-fermion model in a one-dimensional lattice, which consists of a spin-1/2 XX chain coupled to a single itinerant fermion.
In our study, we focus on translationally invariant observables including the charge and energy current, thereby also connecting the ETH with transport properties.
Considering observables with a Hilbert-Schmidt norm of one, we first perform a comprehensive analysis of ETH in the model taking into account latest developments.
A particular emphasis is on the analysis of the structure of the offdiagonal matrix elements $|\langle \alpha | \hat O | \beta \rangle|^2$ in the limit of small eigenstate energy differences $\omega = E_\beta - E_\alpha$.
Removing the dominant exponential suppression of $|\langle \alpha | \hat O | \beta \rangle|^2$, we find that:
(i) the current matrix elements exhibit a system-size dependence that is different from other observables under investigation,
(ii) matrix elements of several other observables exhibit a Drude-like structure with a Lorentzian frequency dependence.
We then show how this information can be extracted from the autocorrelation functions as well.
Finally, our study is complemented by a numerical analysis of the fluctuation-dissipation relation for eigenstates in the bulk of the spectrum.
We identify the regime of $\omega$ in which the well-known fluctuation-dissipation relation is valid with high accuracy for finite systems.
\end{abstract}


\maketitle

\section{Introduction}
\label{sec:intro}

Nonequilibrium experiments with many-body systems in nearly perfect isolation from the environment have become feasible in the last two decades~\cite{greiner02, Kinoshita2006, gring_kuhnert_12, Trotzky2012, langen15a, Schreiber2015, Choi2016, Kaufman2016, Neill2016, tang_kao_18}.
One of the outstanding challenges is to better understand the mechanism of relaxation and thermalization in generic quantum systems, i.e., systems that do not exhibit any nontrivial local conservation laws.
In particular, studies of thermalization and its relation to quantum ergodicity now represent a rapidly evolving research field (see, e.g., Refs.~\cite{dalessio_kafri_16,mori_ikeda_18, deutsch_18} for reviews).

An important conceptual goal is to provide a simple framework to explain thermalization in many-body systems on a lattice and away from any perturbative regimes in interaction strengths.
A useful insight can be provided by two related approaches, the random matrix theory (RMT)~\cite{brody_flores_81, mehta_91, dalessio_kafri_16} and quantum typicality~\cite{vonneumann_29, Goldstein2010, popescu_short_06, goldstein_lebowitz_06, reimann_15a, reimann_15b, dymarsky_lashkari_18}. 
For random matrices one can introduce the corresponding RMT ansatz~\cite{dalessio_kafri_16} and use it for matrix elements of observables in Hamiltonian eigenstates $|\alpha\rangle$, $|\beta\rangle$,
\begin{equation} \label{def_rmt_ansatz}
 \langle \alpha | \hat O | \beta \rangle = \frac{\cal N}{\sqrt{\cal D}} R_{\alpha\beta} \,,
\end{equation}
where ${\cal D}$ is the Hilbert-space dimension, ${\cal N} = || \hat O ||$ is the Hilbert-Schmidt norm [see Eq.~(\ref{def_norm})] of the operator $\hat O$, and $R_{\alpha\beta}$ are random numbers with zero mean and a variance of one (two) for the offdiagonal (diagonal) matrix elements.
For simplicity, we set ${\cal N} \equiv 1$ for all operators studied in this paper.
If the RMT ansatz is valid, it provides a stepping stone to guarantee thermalization after a nonequilibrium time evolution under unitary dynamics~\cite{dalessio_kafri_16}.

Nevertheless, the RMT ansatz in Eq.~(\ref{def_rmt_ansatz}) turns out to be a too crude approximation to accommodate all the rich features in the nonequilibrium dynamics governed by physical Hamiltonians.
The main missing aspects are temperature, which gives rise to the characteristic dependence of the diagonal matrix elements of observables on energy, and the observable-specific relaxation dynamics, which is reflected as the observable-specific energy dependence of the underlying offdiagonal matrix elements.
To appreciate the significance of the latter, below we first introduce the ETH and its connection to autocorrelation functions, which is followed by a comprehensive summary of our new results.

\subsection{Eigenstate thermalization hypothesis (ETH)} \label{sec:eth_intro}

A natural but nontrivial extension of the RMT ansatz in Eq.~(\ref{def_rmt_ansatz}) was introduced by Srednicki~\cite{srednicki_99}, now known as the ETH ansatz,
\begin{equation} \label{def_eth_ansatz}
 \langle \alpha | \hat O | \beta \rangle = O(\bar E) \delta_{\alpha\beta} + \frac{1}{\sqrt{e^{S(\bar E)}}} f_O(\bar E, \omega) R_{\alpha\beta} \, .
\end{equation}
In this ansatz, the amplitude of fluctuations $1/\sqrt{e^{S(\bar E)}}$, which replaces $1/\sqrt{\cal D}$ in Eq.~(\ref{def_rmt_ansatz}), is expressed through the thermodynamic entropy $S(\bar E)$ at energy $\bar E$, which scales extensively with the lattice size $L$.
Here, $\bar E = (E_\alpha + E_\beta)/2$ is the average energy of a pair of eigenstates $|\alpha\rangle$ and $|\beta\rangle$ and $\omega=E_\alpha - E_\beta$ is the corresponding energy difference (we set $\hbar \equiv 1$ throughout the exposition).
The structure functions $O(\bar E)$ and $f_O(\bar E,\omega)$ in the ETH ansatz are, in principle, arbitrary smooth functions of their arguments, in contrast to the RMT ansatz where $O(\bar E) = 0$ and $f_O(\bar E, \omega) = 1$.

The ETH~\cite{deutsch_91, srednicki_94, rigol_dunjko_08} provides a sufficient condition for quantum ergodicity: if a system satisfies the ETH, expectation values of observables will time evolve (for the majority of physically relevant initial states) towards the corresponding statistical ensemble average.
Great efforts have been invested to test the validity of the ETH ansatz in finite many-body systems~\cite{rigol_09a, steinigeweg_herbrych_13, khatami_pupillo_13, beugeling_moessner_14, sorg14, steinigeweg_khodja_14, kim_ikeda_14, beugeling_moessner_15, mondaini_fratus_16, mondaini_rigol_17, yoshizawa_iyoda_18, khaymovich_haque_19, jansen_stolpp_19, leblond_mallayya_19, mierzejewski_vidmar_20, brenes_leblond_20, brenes_goold_20, sugimoto_hamazaki_21, richter_dymarsky_20, leblond_rigol_20, noh_21}.
A particular emphasis has been devoted to the verification of the scaling of the amplitude of fluctuations $1/\sqrt{e^{S(\bar E)}}$ and to the study of the degree of randomness encoded in $R_{\alpha\beta}$ in Eq.~(\ref{def_eth_ansatz}).
It is now well established that:
(i) the matrix-element fluctuations decay exponentially with lattice size $L$~\cite{beugeling_moessner_14, kim_ikeda_14, mondaini_fratus_16, yoshizawa_iyoda_18, jansen_stolpp_19, leblond_mallayya_19, mierzejewski_vidmar_20, sugimoto_hamazaki_21, richter_dymarsky_20, leblond_rigol_20, noh_21},
(ii) the distribution of fluctuations is Gaussian~\cite{mukerjee_oganesyan_06, beugeling_moessner_15, leblond_mallayya_19, brenes_leblond_20}, and
(iii) the ratio $\Sigma^2$ of variances of matrix elements (the diagonals divided by the offdiagonals) satisfies the RMT prediction $\Sigma^2 = 2$~\cite{dalessio_kafri_16} in sufficiently narrow energy windows~\cite{mondaini_rigol_17, jansen_stolpp_19, richter_dymarsky_20}.
Examples of (i) and (iii), with remarkable numerical accuracy, are also demonstrated in this work.

Less attention has been devoted to the structure functions $O(\bar E)$ and $f(\bar E, \omega)$.
In the analysis of fluctuations of the diagonal matrix elements, the structure function $O(\bar E)$ is usually subtracted~\cite{beugeling_moessner_14, yoshizawa_iyoda_18}.
Another approach to study fluctuations of the matrix elements is to consider two neighboring eigenstates, which are exponentially close in energy and hence the structure function of the matrix elements does not impact the result~\cite{kim_ikeda_14, mondaini_fratus_16, jansen_stolpp_19, leblond_mallayya_19}.
Recently, a quantitative framework has been introduced to relate the structure of $O(\bar E)$ and the ''similarity'' of observables to local integrals of motion, quantified by the projection of observables onto conserved quantities and products thereof~\cite{mierzejewski_vidmar_20}.
It was argued that, if such similarity of observables to local integrals of motion is removed, then $O(\bar E) \approx 0$, similar to the RMT ansatz in Eq.~(\ref{def_rmt_ansatz}).

Studies of the offdiagonal matrix elements are now also gaining attention~\cite{khatami_pupillo_13, beugeling_moessner_15, dalessio_kafri_16, mondaini_rigol_17, nation_porras_18, dymarsky_liu_19, jansen_stolpp_19, leblond_mallayya_19, brenes_leblond_20, brenes_goold_20, richter_dymarsky_20, leblond_rigol_20, luitz_barlev_16, serbyn_papic_17, sels_polkovnikov_20}, with strong interest in the structure of $|f_O(\bar E,\omega)|^2$ at small $\omega$~\cite{dalessio_kafri_16, luitz_barlev_16, brenes_leblond_20, brenes_goold_20, richter_dymarsky_20, leblond_rigol_20}.
The physical importance of $|f_O(\bar E,\omega)|^2$ is related to the decay of autocorrelation functions and the fine structure of response functions~\cite{dalessio_kafri_16}. 
Moreover, it governs the fluctuation-dissipation theorem for nonequilibrium pure states~\cite{dalessio_kafri_16, kogoj_vidmar_16, noh_sagawa_20} and it determines the heating rates in driven systems~\cite{mallayya_rigol_19}.
Note that there has been interest in the fluctuation-dissipation theorem in nonequilibrium and quench dynamics in closed systems~\cite{srednicki_99, khatami_pupillo_13, nation_porras_19, kogoj_vidmar_16, essler_evangelisti_12, foini_cugliandolo_12, foini_gambassi_17, schuckert_knap_20} from various angles.

Another important question, which we also address here, is the impact of the system size on $|f_O(\bar E,\omega)|^2$.
Extracting and understanding the $L$-dependence of $|f_O(\bar E,\omega)|^2$ is paramount for identifying time scales that define the onset of quantum chaotic behavior. 
In connection with transport, such time scales would relate to the onset of hydrodynamic behavior and generically diffusion.
For instance, in Ref.~\cite{dalessio_kafri_16}, a generalization of the Thouless energy scale to the many-body case was discussed.
There is, however, only a scarce amount of work on concrete models and observables and our work aims at broadening the available information on $L$-dependencies of $|f_O(\bar E,\omega)|^2$~\cite{dalessio_kafri_16, luitz_barlev_16, brenes_leblond_20, brenes_goold_20, richter_dymarsky_20, leblond_rigol_20, dymarsky_18}.

Note that the extraction of the system-size dependence of $|f_O(\bar E,\omega)|^2$, which may strongly depend on a specific observable~\cite{brenes_leblond_20}, is a nontrivial numerical operation.
In particular, while the amplitude of $|f_O(\bar E,\omega)|^2$ at small $\omega$ may scale polynomially with the number of lattice sites $L$~\cite{dalessio_kafri_16}, this effect is masked by the dominant contribution $e^{-S(\bar E)}$ in the ETH ansatz~(\ref{def_eth_ansatz}).
The latter is exponentially small in $L$ since $S(\bar E) \propto L$ for the eigenstates away from the spectral edges.

An increased attention has recently been devoted to the study of the matrix elements of observables in quantum chaotic models in which integrability is broken by a single {\it static} impurity~\cite{santos_04, barisic_prelovsek_09, santos_mitra_11, torresherrera_santos_14, brenes_mascarenhas_18, brenes_leblond_20, brenes_goold_20, santos_perezbernal_20}.
Here, we consider a spin-1/2 XX chain where integrability is broken by a coupling to a single {\it itinerant} impurity (i.e., a single fermion), extending earlier work by some of us on the Holstein-polaron model~\cite{jansen_stolpp_19}.
These types of models are relevant for describing polaron physics~\cite{bonca99, fehske2007}, i.e., properties of itinerant charge carriers coupled to bosonic degrees of freedom.
We study different classes of observables and discuss the underlying physical features that can be identified from their offdiagonal matrix elements.

\subsection{Autocorrelation functions and spectral densities} \label{sec:auto_intro}

A convenient approach to extract properties of the structure function $|f_O(\bar E,\omega)|^2$ is provided by the autocorrelation functions and the corresponding spectral densities of the observable $\hat O$.
One possibility is to consider local observables of the conserved quantities, which in quantum chaotic systems should propagate diffusively~\cite{bertini_heidrichmeisner_21, luitz_barlev_17}, giving rise to a specific low-$\omega$ form of the structure function~\cite{dalessio_kafri_16}.
Here, in contrast, we focus on various translationally invariant observables and study scaling properties of their structure functions $|f_O(\bar E,\omega)|^2$.

We define the symmetric autocorrelation function for an observable $\hat O$ in an eigenstate $|\alpha\rangle$ as
\begin{equation} \label{def_C_alpha}
 C_O^{(\alpha)} (t) = \langle \alpha | \{ \hat O(t) \hat O(0) \} | \alpha \rangle_{\rm c} \,,
\end{equation}
where $\hat O(t) = e^{i \hat H t} \hat O e^{-i \hat H t}$ and $\{ \, \cdots \, \}$ denotes the anticommutator.
For convenience, the function in Eq.~(\ref{def_C_alpha}) is {\it connected}, implying that for every pair of operators $\hat O_1$ and $\hat O_2$ we have
$\langle \hat O_1 \hat O_2 \rangle_{\rm c} \equiv \langle \hat O_1 \hat O_2 \rangle - \langle \hat O_1 \rangle \, \langle \hat O_2 \rangle$.
One can express Eq.~(\ref{def_C_alpha}) in terms of the matrix elements of observables as
\begin{equation} \label{def_C_alpha_matele}
 C_O^{(\alpha)} (t) = 2 \sum_{\beta \neq \alpha} \cos\left[(E_\beta-E_\alpha) t\right] \, |\langle \alpha | \hat O | \beta \rangle|^2 \,,
\end{equation}
where the sum runs over all eigenstates $\{ |\beta\rangle \}$ in the Hilbert space except for $|\beta\rangle = |\alpha\rangle$, which cancels out for the connected autocorrelation functions.

A Fourier transform of the autocorrelation function $C_O^{(\alpha)}(t)$ is referred to as the spectral density $S_{O,+}^{(\alpha)}(\omega)$ of an operator $\hat O$,
\begin{equation} \label{def_S_alpha}
 S_{O,+}^{(\alpha)}(\omega) = \int_{-\infty}^\infty dt \, e^{i \omega t} \, C_O^{(\alpha)}(t) \,,
\end{equation}
where the subindex ''+'' in $S(\omega)$ denotes the use of the anticommutator in Eq.~(\ref{def_C_alpha}).
Applying the ETH ansatz~(\ref{def_eth_ansatz}) for the matrix elements of observables, one can show (see Sec.~\ref{sec:spectrum} or~\cite{dalessio_kafri_16} for the derivation) that the leading term of the spectral density equals
\begin{equation} \label{def_S_alpha_short}
 S_{O,+}^{(\alpha)}(\omega) = 4\pi \cosh(\beta \omega/2) |f_O(E_\alpha, \omega)|^2 \,.
\end{equation}
If the autocorrelation function in Eq.~(\ref{def_C_alpha}) is defined by the commutator, it is proportional to the response functions, whose spectral density $S_{O,-}^{(\alpha)}(\omega)$ can be written in a similar form [see Eq.~(\ref{def_S_alpha_minus_short})].
The relation between $S_{O,+}^{(\alpha)}(\omega)$ and $S_{O,-}^{(\alpha)}(\omega)$ constitutes the basis of the eigenstate fluctuation-dissipation theorem~\cite{dalessio_kafri_16}.
However, in finite systems, their relation may be governed by subleading terms, which include the derivatives of $|f_O(E_\alpha, \omega)|^2$~\cite{dalessio_kafri_16, noh_sagawa_20}.
In this work, we identify the regime of $\omega$ in which the fluctuation-dissipation theorem, without the inclusion of subleading terms, is valid with high accuracy in finite systems.

Another important quantity is the time integral of the autocorrelation function in Eq.~(\ref{def_C_alpha}), which can be calculated numerically using various techniques such as the exact time evolution~\cite{prosen_99, steinigeweg_gemmer_09}, matrix-product state methods~\cite{karrasch_bardarson_12, paeckel_koehler_19}, dynamical quantum typicality~\cite{stenigeweg_gemmer_14, stenigeweg_heidrichmeisner_14}, and the numerical linked cluster expansion~\cite{richter_steinigeweg_19, mallayya_rigol_18, guardadosanchez_brown_18, white_sundar_17}.
At sufficiently long times, this quantity reveals the $\omega\to 0$ properties of $|f_O(\bar E, \omega)|^2$.
For specific operators in quantum chaotic models, such as the currents, this quantity is expected to become a constant which has a well-defined physical meaning, namely, it is the diffusion constant~\cite{steinigeweg_gemmer_09, bertini_heidrichmeisner_21}.
The latter property relates the low-energy structure of $|f_O(\bar E, \omega)|^2$ to transport~\cite{bertini_heidrichmeisner_21}.
Here, we contrast the structure function $|f_O(\bar E, \omega)|^2$ of currents with those of other observables, and we explicitly demonstrate how the time integral of the autocorrelation functions for currents approaches results obtained by averaging the matrix elements in the $\omega\to 0$ limit.

\subsection{Summary of results} \label{sec:sum}

In this work, we study a nonintegrable spin-fermion model introduced in Sec.~\ref{sec:model}, which consists of the spin-1/2 XX chain and a single itinerant fermion.
In this model, integrability is broken by a local coupling of spins to the fermion.
The model exhibits two conserved quantities, the number of fermions and the total energy, and it is equivalent to the Holstein-polaron model with dispersive phonons subject to a hard-core constraint.
Apart from the currents, we also study various other observables defined in Sec.~\ref{sec:observables}.

In the first part of the paper [cf.~Secs.~\ref{sec:diagonals}-\ref{sec:variances}], we scrutinize many aspects of the ETH ansatz in Eq.~(\ref{def_eth_ansatz}).
Specifically, we study the diagonal matrix elements in Sec.~\ref{sec:diagonals}, the offdiagonal matrix elements in Sec.~\ref{sec:structure} and the corresponding variances of matrix elements in Sec.~\ref{sec:variances}.
There are two main results of this analysis:
(i) We identify two classes of observables.
The first class are the current operators, for which the scaled offdiagonal matrix elements $\overline{|O_{\alpha\beta}|^2} {\cal D}$ from the bulk of the spectrum exhibit almost no system-size dependence.
(All other observables under investigation exhibit a robust system size dependence in the low-$\omega$ regime.)
The second class are the operators for which the offdiagonal matrix elements $\overline{|O_{\alpha\beta}|^2}$ exhibit a Drude-like (Lorentzian) dependence on $\omega$.
(ii) For those observables whose diagonal matrix elements exhibit no structure in the bulk of the spectrum [i.e., $O(\bar E) \approx 0$ in Eq.~(\ref{def_eth_ansatz})], we verify the ETH ansatz with remarkable accuracy:
the variance of the offdiagonal matrix elements scales as $a_0 {\cal D}^{-\gamma}$, where the numerical values of $a_0$ and $\gamma$ equal 1 on two digits.
Moreover, the ratio of variances (diagonal versus the offdiagonal) equals 2 on nearly three digits.
These results provide simple and powerful benchmarks for future studies.

In the second part [cf.~Secs.~\ref{sec:spectrum}-\ref{sec:autocorrelation}], we focus on properties of the offdiagonal matrix elements as extracted from the autocorrelation functions and the spectral densities.
We first study the validity of the fluctuation-dissipation theorem for eigenstates in the bulk of the spectrum.
In particular, we identify the regime of energies in which the fluctuation-dissipation relation, valid in the thermodynamic limit, holds with high accuracy in finite systems accessible with exact diagonalization.
We then study the time dependence of the autocorrelation functions, or the integrals thereof, for various observables.
This allows us to make explicit connections between properties of the time-evolving quantities and the specific features of the offdiagonal matrix elements of observables.
We conclude in Sec.~\ref{sec:conclusion}.

\section{Model}
\label{sec:model}

We study a spin-fermion Hamiltonian on a one-dimensional lattice with $L$ sites, which consists of the spin-1/2 XX chain in a homogeneous magnetic field, coupled to a single itinerant spinless fermion.
The Hamiltonian is given by
\begin{align} \label{def_H}
 \hat H = & - t_0 \sum_{j=1}^L \left( \hat c_j^{\dag} \hat c_{j+1}^{\phantom{\dag}} + \hat c_{j+1}^{\dag} \hat c_j^{\phantom{\dag}} \right) 
 + g \sum_{j=1}^L \hat n_j \hat S_j^x \nonumber \\
 & + \omega_0 \sum_{j=1}^L \hat S_j^z + J \sum_{j=1}^L \left( \hat S_j^x \hat S_{j+1}^x + \hat S_j^y \hat S_{j+1}^y \right) \, ,
\end{align}
where $\hat c_j$ is a spinless fermion annihilation operator at site $j$, $\hat n_j = \hat c_j^\dagger \hat c_j$ is the fermion site-occupation operator, and $\hat S_j^\alpha$ (for $\alpha = \{x,y,z\}$) are standard spin-1/2 operators.
We use periodic boundary conditions, $\hat c_{L+1} \equiv \hat c_1$ and $\hat S_{L+1}^\alpha \equiv \hat S_1^\alpha$.
We set the unit of energy to $t_0 \equiv 1$ and the lattice spacing to $a \equiv 1$.

The Hamiltonian in Eq.~(\ref{def_H}) conserves the total number of fermions $\hat N = \sum_j \hat c_j^\dagger \hat c_j$, but not the total spin magnetization $\hat S^z = \sum_j \hat S_j^z$.
Here, we focus on the single fermion sector $\langle \hat N \rangle = 1$, for which the Hilbert-space dimension is ${\cal \tilde D} = L \, 2^L$.
Using full exact diagonalization and employing translational invariance, we numerically calculate all eigenvalues $\{ E_\alpha \}$ and eigenstates $\{ |\alpha\rangle \}$ of the Hamiltonian in Eq.~(\ref{def_H}) in a target total quasimomentum sector $k$ with the Hilbert-space dimension ${\cal D} = 2^L $.
We focus on $k=2\pi/L$ throughout the work.
Unless stated otherwise, we set the parameters to $\omega_0/t_0=1/2$, $g/t_0=\sqrt{2}$ and $J/t_0 = 0.2$.

The spin-fermion Hamiltonian in Eq.~(\ref{def_H}) is equivalent to the Holstein-polaron model with hard-core bosons, 
\begin{align} \label{def_Holstein}
 \hat H = & - t_0 \sum_{j=1}^L \left( \hat c_j^{\dag} \hat c_{j+1}^{\phantom{\dag}} + \hat c_{j+1}^{\dag} \hat c_j^{\phantom{\dag}} \right) 
 + \gamma' \sum_{j=1}^L \hat n_j (\hat b_j^\dagger + \hat b_j) \nonumber \\
 & + \omega_0 \sum_{j=1}^L \hat b_j^\dagger \hat b_j + \omega' \sum_{j=1}^L \left( \hat b_j^\dagger \hat b_{j+1} + \hat b_{j+1}^\dagger \hat b_j \right) \, ,
\end{align}
where $\hat b_j$ is a boson annihilation operator on site $j$, and infinite repulsion is enforced by the constraints $(\hat b_j)^2 = (\hat b_j^\dagger)^2 = 0$.
The Hamiltonian in Eq.~(\ref{def_Holstein}) has recently been investigated in studies of quantum ergodicity and nonequilibrium dynamics~\cite{jansen_stolpp_19, kogoj_mierzejewski_16}, as well as in studies of response functions~\cite{bonca_20}.
The spectra of Hamiltonians in Eqs.~(\ref{def_H}) and~(\ref{def_Holstein}) are identical (up to a constant energy shift) if $g = 2\gamma'$ and $J = 2\omega'$.
Note that a nonzero $\omega'$ implies that the phonons are dispersive, which is usually not taken into account in studies of the Holstein-polaron model.

In Ref.~\cite{jansen_stolpp_19}, it has been shown that the Holstein-polaron model with a hardcore constraint for the phonons [equivalent to  Eq.~(\ref{def_Holstein})] is quantum chaotic in the sense that the level statistics at $\omega'=J=0$ agrees with the one predicted by the Gaussian orthogonal ensemble in a wide range of couplings $\gamma'$.
Here, we work with the spin-fermion Hamiltonian representation and focus on properties of $\hat H$ in Eq.~(\ref{def_H}) at nonzero $J$.

\section{Set of Observables}
\label{sec:observables}

We study nine different dimensionless observables defined below.
All observables are traceless operators,
\begin{equation} \label{def_traceless}
 \langle \hat O \rangle = {\cal D}^{-1} {\rm Tr}\{ \hat O \} = {\cal D}^{-1} \sum_{\alpha=1}^{\cal D} \langle\alpha|\hat O |\alpha\rangle = 0 \, ,
\end{equation}
which are normalized as $|| \hat O || = 1$.
The norm is defined by the Hilbert-Schmidt norm (also denoted as the Frobenius norm)
\begin{equation} \label{def_norm}
 || \hat O ||^2 = {\cal D}^{-1} {\rm Tr}\{ \hat O^2 \} = {\cal D}^{-1} \sum_{\alpha, \beta = 1}^{\cal D} |\langle\alpha|\hat O |\beta\rangle|^2 = 1 \, .
\end{equation}
Throughout the paper, we use a compact notation for the matrix elements of observables $O_{\alpha\beta} \equiv \langle \alpha | \hat O | \beta \rangle$.

\subsection{Currents}

A particular focus of this work is on current operators.
They are related to the conserved U(1) quantities of the system through the continuity equation.
For the Hamiltonian~(\ref{def_H}) under investigation, there exist two such conserved quantities: the total number of fermions $\hat N$ and the total energy $\hat H$.
They are connected to the charge current $\hat J_{\rm c}$ and the energy current $\hat J_{\rm e}$, respectively.
The charge current is defined as
\begin{equation} \label{def_Jc}
 \hat J_{\rm c} = \sum_{j=1}^L \left( i\hat c_j^\dagger \hat c_{j+1} - i\hat c_{j+1}^\dagger \hat c_{j} \right) \,,
\end{equation}
and the energy current is defined as
\begin{align}
 \hat J_{\rm e} = \sum_{j=1}^L \bigg( & \hat \jmath_{j}^{(2)} - \frac{g}{t_0} \, \left( \hat \jmath_{j}^{(1)} \hat S_{j+1}^x \right) - \frac{gJ}{2t_0^2} \left( \hat n_{j+1} \hat S_j^y \right) \nonumber \\
 + & \frac{w_0 J}{t_0^2} \left(\hat S_{j}^x \hat S_{j+1}^y - \hat S_{j}^y \hat S_{j+1}^x\right) \\
 + & \frac{J^2}{2t_0^2} \left(\hat S_j^x \hat S_{j+2}^y - \hat S_j^y \hat S_{j+2}^x\right)
 \bigg) \,, \nonumber
\end{align}
where $\hat \jmath_j^{(m)} = i \hat c_{j}^\dagger \hat c_{j+m} - i \hat c_{j+m}^\dagger \hat c_{j}$ is a generalized current operator that moves particles a distance $m$.

Since the operators are block diagonal in the basis of translationally invariant states used for numerical diagonalization, we make them traceless and normalized, to satisfy Eqs.~(\ref{def_traceless}) and~(\ref{def_norm}), within a target symmetry sector (i.e., the single fermion sector with total quasimomentum $k$).
To this end, we renormalize the currents in these symmetry sectors as
\begin{equation}
 \hat J \;\; \to \;\; {\cal N} \, \hat J + {\rm const}\,,
\end{equation}
where for the charge current, we get ${\cal N} = 1/\sqrt{2-\epsilon_\rho}$, with $\epsilon_\rho = \cos(2k)/2^{L-3+(L\,{\rm mod}\,2)} + \sin^2(k)/2^{2L-4}$, and ${\rm const} = {\cal N} \sum_j \rho_0$, with $\rho_0 = \sin(k)/2^{L-2}$.
For the energy currents, we calculate both normalization parameters numerically.

\begin{figure*}[!]
\includegraphics[width=2.0\columnwidth]{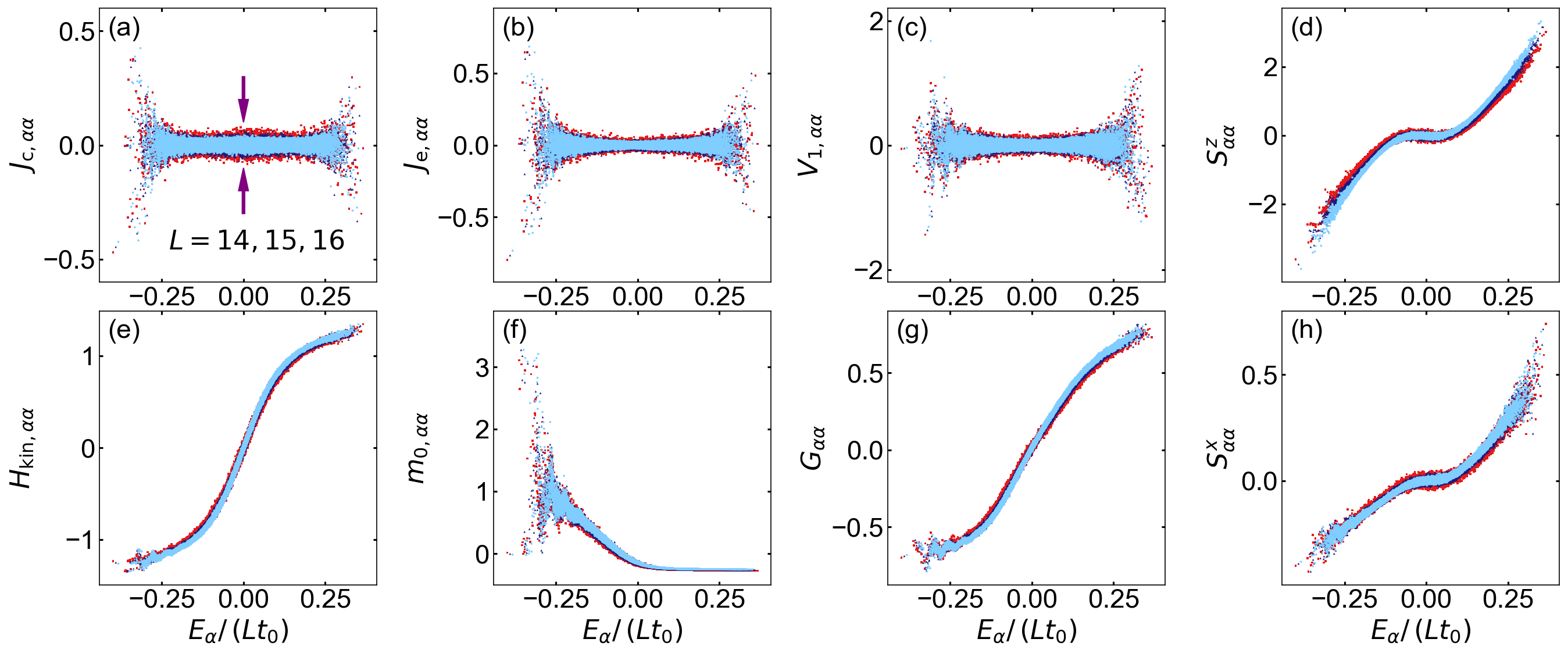}
\caption{
Diagonal matrix elements $O_{\alpha\alpha}$ of the eight observables defined in Sec.~\ref{sec:observables}, plotted as a function of $E_\alpha/(L t_0)$.
Symbols from the back to the front represent results for $L=14$ (red), $L=15$ (dark blue) and $L=16$ (light blue), respectively.
}
\label{fig:eev1}
\end{figure*}

\begin{figure}[!]
\includegraphics[width=0.90\columnwidth]{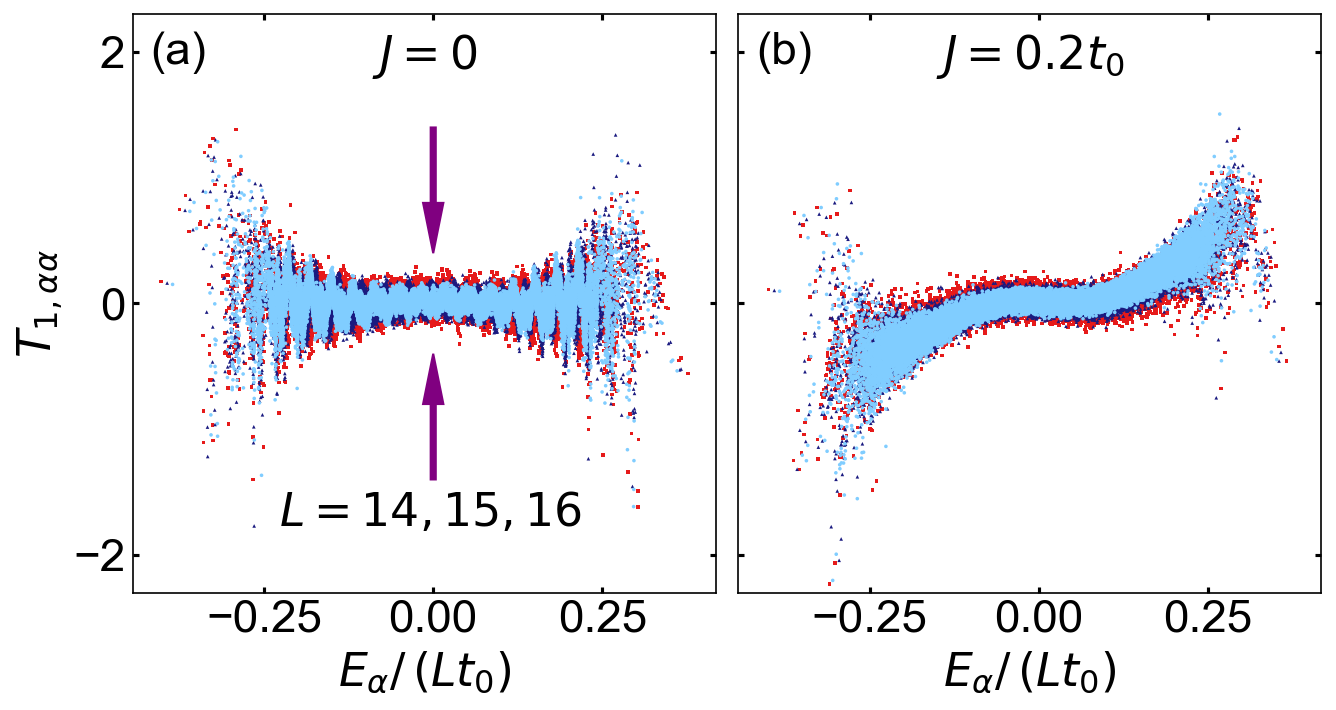}
\caption{
Diagonal matrix elements $T_{1,\alpha\alpha}$ of the spin correlator $\hat T_1$ defined in Eq.~(\ref{def_T1}).
Results are plotted as a function of $E_\alpha/(L t_0)$ for three system sizes $L=14$ (red), $L=15$ (dark blue) and $L=16$ (light blue).
The spin-exchange coupling is set to (a) $J = 0$ and (b) $J/t_0 = 0.2$.
}
\label{fig:eev2}
\end{figure}

\subsection{Additional translationally invariant observables}

In addition to the currents, we study several other normalized observables that are extensive sums of local operators with support on at most two consecutive lattice sites [with the exception of the fermion quasimomentum distribution, to be defined in Eq.~(\ref{def_mq})].
Among the operators that act only on the spin sector, we study the spin correlators
\begin{equation} \label{def_T1}
 \hat T_1 = \frac{2\sqrt{2}}{\sqrt{L}} \sum_{j=1}^L \left(\hat S_j^x \hat S_{j+1}^x + \hat S_{j}^y \hat S_{j+1}^y \right) \, 
 \end{equation}
and
\begin{equation}
 \hat V_1 =  \frac{2\sqrt{2}}{\sqrt{L}} \sum_{j=1}^L \left( \hat S_j^x \hat S_{j+1}^y - \hat S_{j}^y \hat S_{j+1}^x \right) \, ,
\end{equation}
as well as the total spin polarization along the $z$-axis,
\begin{equation}
 \hat S^z = \frac{2}{\sqrt{L}} \sum_{j=1}^L \hat S_j^z \,,
\end{equation}
and along the $x$-axis,
\begin{equation} \label{def_x}
 \hat S^x = \frac{2}{\sqrt{L}} \sum_{j=1}^L \hat S_j^x\, .
\end{equation}
We also study the spin-fermion coupling
\begin{equation}
 \hat G = 2 \sum_{j=1}^L \hat n_j \hat S_j^x \,.
\end{equation}
Among the operators that act only on the fermion sector, we study the fermion kinetic energy
\begin{equation}
 \hat H_{\rm kin} = - \sum_{j=1}^L \left( \hat c_j^\dagger \hat c_{j+1} + \hat c_{j+1}^\dagger \hat c_{j} \right) \,
\end{equation}
and the fermion quasimomentum distribution
\begin{equation} \label{def_mq}
 \hat m_q = \sum_{j,l=1}^L e^{i(l-j)q} \hat c_j^\dagger \hat c_l \,.
\end{equation}
For the latter, we study the $q=0$ contribution only.
The operator $\hat m_q$ is a one-body but nonlocal observable.

In analogy to the currents, we make $\hat H_{\rm kin}$ and $\hat m_q$ traceless and normalized within the target quasimomentum sector.
To this end, we renormalize $\hat H_{\rm kin} \to {\cal N}_\kappa \hat H_{\rm kin} - {\cal N}_\kappa \sum_j \kappa_0$, where ${\cal N}_\kappa = 1/\sqrt{2 + \epsilon_\kappa}$ and the constants $\epsilon_\kappa$ and $\kappa_0$ are exponentially small in $L$, namely, $\epsilon_\kappa = \cos(2k)/2^{L-3+(L\,{\rm mod}\,2)} - \cos^2(k)/2^{2L-4}$ and $\kappa_0 = \cos(k)/2^{L-2}$.
We also renormalize $\hat m_q \to {\cal N}_q \hat m_q + {\rm const.}$, where the normalization constants are calculated numerically.

\section{Structure of diagonal matrix elements} \label{sec:diagonals}

The diagonal matrix elements $O_{\alpha\alpha}$ of all observables defined in Sec.~\ref{sec:observables} are shown in Fig.~\ref{fig:eev1} and Fig.~\ref{fig:eev2}.
A few comments can be made about their properties.

First, all the diagonal matrix elements appear to be consistent with the ETH, i.e., they are smooth functions of energy, up to fluctuations.
A quantitative study of their fluctuations is carried out in Sec.~\ref{sec:variances}, where we calculate variances of the matrix elements.

Second, the diagonal matrix elements of some observables [cf.~$\hat J_{\rm c}$, $\hat J_{\rm e}$ and $\hat V_1$ in Figs.~\ref{fig:eev1}(a)-\ref{fig:eev1}(c), respectively] appear to be structureless, i.e., $O(\bar E)\approx 0$, since their matrix elements fluctuate around zero at all energies (away from spectral edges).
This feature seems to be in agreement with the RMT ansatz for matrix elements in Eq.~(\ref{def_rmt_ansatz}) in which $O(\bar E) = 0$.
The origin of such a behavior was studied in Ref.~\cite{mierzejewski_vidmar_20}, where it was argued that the degree of structure of the diagonal matrix elements in nonintegrable systems can be quantified by the projection of an observable onto the Hamiltonian $\hat H$ and onto higher powers of $\hat H$.
In particular, in the case of a vanishing projection $\langle \hat O \hat H^n \rangle \equiv {\cal D}^{-1} \sum_\alpha \langle \alpha | \hat O \hat H^n | \alpha \rangle \approx 0$ for all $n$, the matrix elements should exhibit no structure, which appears to be consistent with the results presented in Figs.~\ref{fig:eev1}(a)-\ref{fig:eev1}(c).
In contrast, observables with considerable structure are those which are parts of the Hamiltonian since in such cases, $\langle \hat O \hat H^n \rangle \neq 0$ already for $n=1$.
This is the case for the observables $\hat S^z$, $\hat H_{\rm kin}$ and $\hat G$ shown in Figs.~\ref{fig:eev1}(d), \ref{fig:eev1}(e), and~\ref{fig:eev1}(g), respectively.

Finally, we note that the structure of the diagonal matrix elements for the observables under investigation may strongly depend on parameters of the Hamiltonian.
This is illustrated in Fig.~\ref{fig:eev2} that   shows that the observable $\hat T_1$ is (to a large degree) structureless if $J=0$ [see Fig.~\ref{fig:eev2}(a)]. $\hat T_1$ becomes part of the Hamiltonian~(\ref{def_H}) for $J \neq 0$, implying $\langle \hat T_1 \hat H \rangle \neq 0$ and then,  the structure of its matrix elements is considerably modified [see Fig.~\ref{fig:eev2}(b)].

\section{Structure of offdiagonal matrix elements}
\label{sec:structure}

Next, we study the structure of the offdiagonal matrix elements.
We average the matrix elements for each target mean energy $\bar E_{\rm tar}$ and energy difference $\omega_{\rm tar}$ in a narrow energy window, such that for all matrix elements included in the average, $\bar E = (E_\alpha+E_\beta)/2 \approx \bar E_{\rm tar}$ and $\omega = E_\beta - E_\alpha \approx \omega_{\rm tar}$.
Specifically, we define the average as
\begin{equation} \label{def_structure_offdiag}
\overline{|O_{\alpha\beta}|^2} = \frac{1}{\cal M} \sum_{\substack{\alpha', \beta'; \, \alpha' \neq \beta' \\
|(E_{\alpha'} + E_{\beta'}) / 2 - \bar E_{\rm tar}|<  \Delta/2 \\
||E_{\alpha'} - E_{\beta'}| - \omega_{\rm tar}| < \delta \omega}} |O_{\alpha'\beta'}|^2 \,,
\end{equation}
where $\cal M$ is the number of elements in the sum.
Unless stated otherwise, we set $\bar E_{\rm tar} = 0$, $\Delta/L = 0.0025t_0$ and we choose $\delta \omega$ such that ${\cal M} = 700, 1000, 2000, 3000, 4000$ for $L=12,13,14,15,16$, respectively.
We denote $\bar E_{\rm tar} \to \bar E$ and $\omega_{\rm tar} \to \omega$ further on.

We study the dependence of $\overline{|O_{\alpha\beta}|^2}$ on $\omega$ for different observables in Figs.~\ref{fig:offdiag_current}-\ref{fig:offdiag_large_w}.
A particular attention is devoted to the properties of the scaled matrix elements $\overline{|O_{\alpha\beta}|^2} {\cal D}$.
Such definition of scaled matrix elements is particularly convenient for the normalized operators studied here, since the value of the typical scaled matrix element is $\overline{|O_{\alpha\beta}|^2} {\cal D} \approx 1$.
One may nevertheless argue that a proper rescaling, consistent with the ETH ansatz~(\ref{def_eth_ansatz}), should include the density of states $\rho$ instead of ${\cal D}$, as done in the derivation of the fluctuation-dissipation theorem in Sec.~\ref{sec:spectrum}, see, e.g.,~Eq.~(\ref{expand_rho}).
The latter is proportional to ${\cal D}/\Gamma$, where $\Gamma$ is the width of the density of states, $\Gamma^2 = {\rm Tr}\{\hat H^2\}/{\cal D} - ({\rm Tr}\{\hat H\}/{\cal D})^2$.
For the Hamiltonian in Eq.~(\ref{def_H}), we get
\begin{equation}
\Gamma^2 = 2 t_0^2 + \left(\frac{g}{2}\right)^2 + L \left( \frac{\omega_0^2}{4} + \frac{J^2}{8} \right) \,.
\end{equation}
The $L$ independent part is the contribution from the itinerant impurity, while the $L$ dependent part is the contribution from the extensive spin-1/2 chain.
For the model parameters under investigation [see the text below Eq.~(\ref{def_H})], and for the given system sizes $L$, the first contribution is dominant, and hence $\Gamma$ exhibits only a weak $L$ dependence.
We therefore omit $\Gamma$ in the calculation of the scaled matrix elements.
It should be mentioned, however, that the scaling properties of $\Gamma$ may impact results for system sizes beyond those available from current exact diagonalization studies.
We therefore refrain from making statements about the $L$ dependence of the scaled matrix elements $\overline{|O_{\alpha\beta}|^2} {\cal D}$ in the thermodynamic limit.

We focus on the low-$\omega$ properties, which govern the long-time behavior of the autocorrelation functions.
It is expected that for sufficiently small $\omega \lesssim \Omega_O$, the ETH ansatz~(\ref{def_eth_ansatz}) resembles the RMT ansatz~(\ref{def_rmt_ansatz}) in the sense that
$f_O(\bar E, \omega)$ is independent of $\omega$~\cite{dalessio_kafri_16}.
For moderately large values  of $\Omega_O \lesssim \omega $ but not yet in the regime of a Gaussian decay at $\omega_0/t_0 \gg 1$, the structure function $|f_O(\bar E, \omega)|^2$, or the scaled matrix elements $\overline{|O_{\alpha\beta}|^2} {\cal D}$, exhibit some observable-dependent behavior, which eventually decays towards zero for very large $\omega$.
This regime of intermediate values of $\omega$ determines how the autocorrelation functions in quantum chaotic systems decay towards zero.
We find  that in terms of the system-size dependence of $\overline{|O_{\alpha\beta}|^2} {\cal D}$, the observables can be divided into two classes: the currents and the other observables.

\subsection{Charge and energy current} \label{sec:current}

\begin{figure}[!]
\includegraphics[width=1.0\columnwidth]{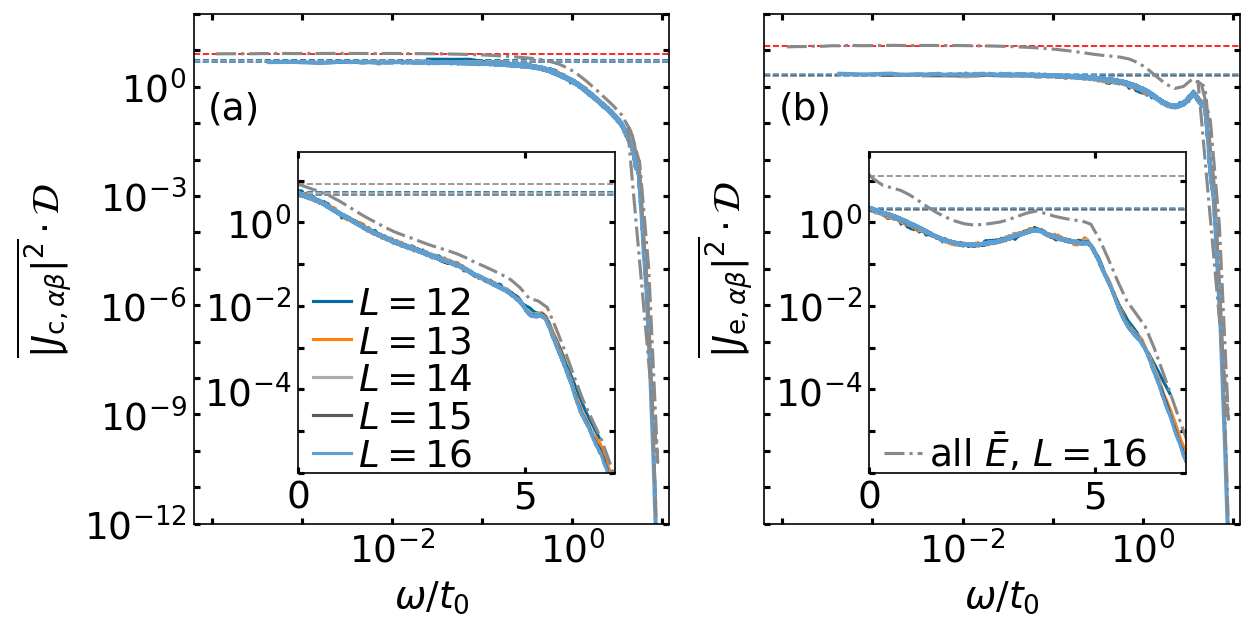}
\caption{
Scaled offdiagonal matrix elements $\overline{|O_{\alpha\beta}|^2} {\cal D}$  for (a) the charge current $\hat J_{\rm c}$ and (b) the energy current $\hat J_{\rm e}$, shown as a function of $\omega/t_0$ in a logarithmic scale (main panels) and in a linear scale (insets).
Overlapping solid lines are results for different system sizes from $L=12$ to $L=16$ at the target energy $\bar E_{\rm tar}=0$ as explained in the context of Eq.~(\ref{def_structure_offdiag}).
Dashed-dotted lines are results for $L=16$, averaged over all energies [$\Delta \to \infty$ in Eq.~(\ref{def_structure_offdiag})].
Horizontal lines are fits of a constant to the data at $\omega/t_0 \ll 1$.
}
\label{fig:offdiag_current}
\end{figure}

We first study the offdiagonal matrix elements of the charge and energy currents.
Their scaled matrix elements $\overline{|O_{\alpha\beta}|^2} {\cal D}$ are shown in Fig.~\ref{fig:offdiag_current}.
A generic property of both currents is the existence of a plateau (i.e., an $\omega$-independent regime) in the limit $\omega/t_0 \to 0$, which can be observed in a log-log scale in the main panels of Fig.~\ref{fig:offdiag_current}.
For the model parameters under investigation, the width $\Omega_O$ of the plateau is roughly $\Omega_{O}/t_0 \approx 10^{-1}$.
Note, however, that the precise value of $\Omega_{O}/t_0$ depends on the number of matrix elements included in the average in Eq.~(\ref{def_structure_offdiag}), as seen from the differences between the solid and dashed-dotted lines in the main panels of Fig.~\ref{fig:offdiag_current}.

In contrast to the $\omega/t_0 \ll 1$ limit, the regime at moderate energies $1\lesssim \omega/t_0 \lesssim 5$ exhibits nongeneric properties, as shown in the insets of Fig.~\ref{fig:offdiag_current}.
While the matrix elements decay nearly exponentially with $\omega$ for the charge current, they exhibit a cusp at $\omega/t_0 \approx 4$ for the energy current.
The nearly exponential decay of $\overline{|O_{\alpha\beta}|^2} {\cal D}$ for the charge current may share some similarities with an anomalous dependence of the optical conductivity on $\omega$ in some other nonintegrable models at high temperatures, such as the extended $t$-$V$ model in one dimension~\cite{mukerjee_oganesyan_06} and strongly correlated models in two dimensions~\cite{jaklic_prelovsek_00, ulaga_mravlje_21}.
The decay of matrix elements at even larger $\omega/t_0 \gtrsim 5$ is studied in Fig.~\ref{fig:offdiag_large_w} in Sec.~\ref{sec:large_w}.

Perhaps the most intriguing feature of the scaled matrix elements $\overline{|O_{\alpha\beta}|^2} {\cal D}$ for the currents is the absence of any pronounced system-size dependence.
This can be seen from (i) the nearly perfect overlap of results for different lattice sizes $L$ in Fig.~\ref{fig:offdiag_current}, and (ii) the analysis in Fig.~\ref{fig:offdiag_plateau_hight}(b) where the values of the scaled matrix elements as a function of $L$ and in the $\omega/t_0 \to 0$ limit turn out to be roughly a constant.
In what follows, we argue that all other observables under investigation exhibit a robust $L$ dependence of the scaled matrix elements $\overline{|O_{\alpha\beta}|^2} {\cal D}$ in the $\omega/t_0 \to 0$ limit.
We note that the $L$ dependence of the scaled matrix elements $\overline{|O_{\alpha\beta}|^2} {\cal D}$ is typically polynomial with $L$, and can only be observed after the dominant, exponential dependence on $L$ is removed by multiplying the matrix elements with ${\cal D}$.
These observations suggest that, at least for the model under investigation, the matrix elements of the currents exhibit a unique scaling property.
Recently, special scaling properties of the particle-current operators (despite being different from those reported here) were studied in the context of ballistic transport in Heisenberg chains, in which integrability is broken by a single static impurity~\cite{brenes_leblond_20}.

In Sec.~\ref{sec:autocorrelation}, we show that certain features of the system-independence of $\overline{|O_{\alpha\beta}|^2} {\cal D}$ can also be detected from the current autocorrelation functions.
Specifically, since the width of the plateau of the offdiagonal matrix-element structure function appears to be $L$ independent, the integral over the autocorrelation function becomes time independent after a moderately short time of the order $(\Omega_O/t_0)^{-1}$.

\subsection{Drude-like structure of matrix elements} \label{sec:drude}

\begin{figure}[!]
\includegraphics[width=1.0\columnwidth]{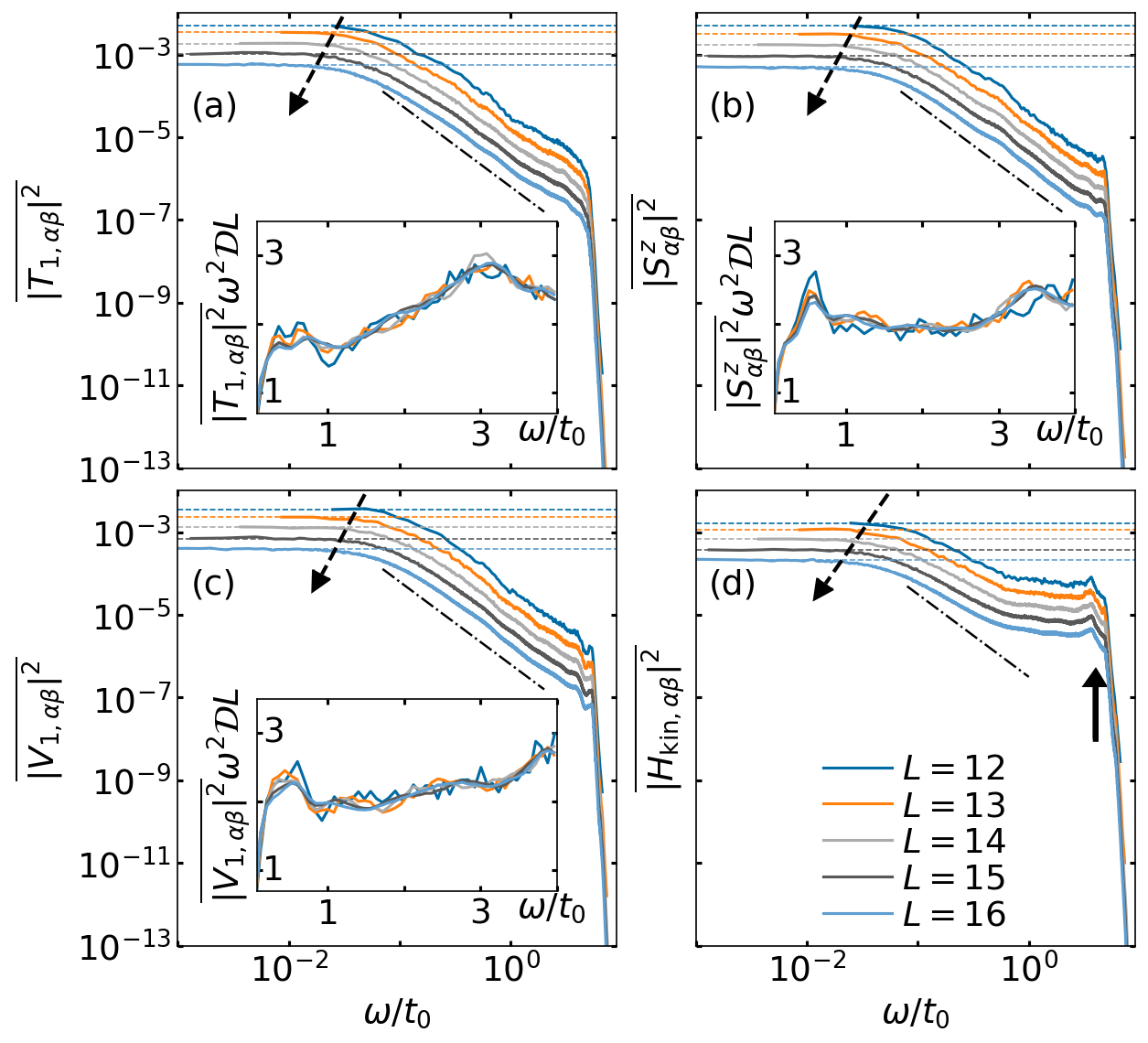}
\caption{
Offdiagonal matrix elements $\overline{|O_{\alpha\beta}|^2}$ as a function of $\omega/t_0$ for the observables (a) $\hat T_1$, (b) $\hat S^z$, (c) $\hat V_1$, and (d) $\hat H_{\rm kin}$.
Results are shown for different lattice sizes from $L=12$ to $L=16$.
Dashed horizontal lines are fits of the data at $\omega/t_0 \ll 1$ to a constant.
Dashed-dotted lines denote the scaling $\propto (\omega/t_0)^{-2}$.
The arrow in (d) marks the peak at $\omega/t_0 \approx 4$.
Insets show the scaled matrix elements $\overline{|O_{\alpha\beta}|^2} \omega^2 {\cal D}L$.
In the insets, we choose $\delta \omega$ in Eq.~(\ref{def_structure_offdiag}) as $\delta\omega/t_0 = 0.08$.
}
\label{fig:offdiag_1w2}
\end{figure}

Next, we focus on the offdiagonal matrix elements of the observables $\hat T_1$, $\hat V_1$, $\hat S^z$, and $\hat H_{\rm kin}$.
The choice of these observables is based on some common features that they exhibit in the structure of the matrix elements and/or in the autocorrelation functions studied in Sec.~\ref{sec:autocorrelation}.

\begin{figure*}[!]
\includegraphics[width=2.0\columnwidth]{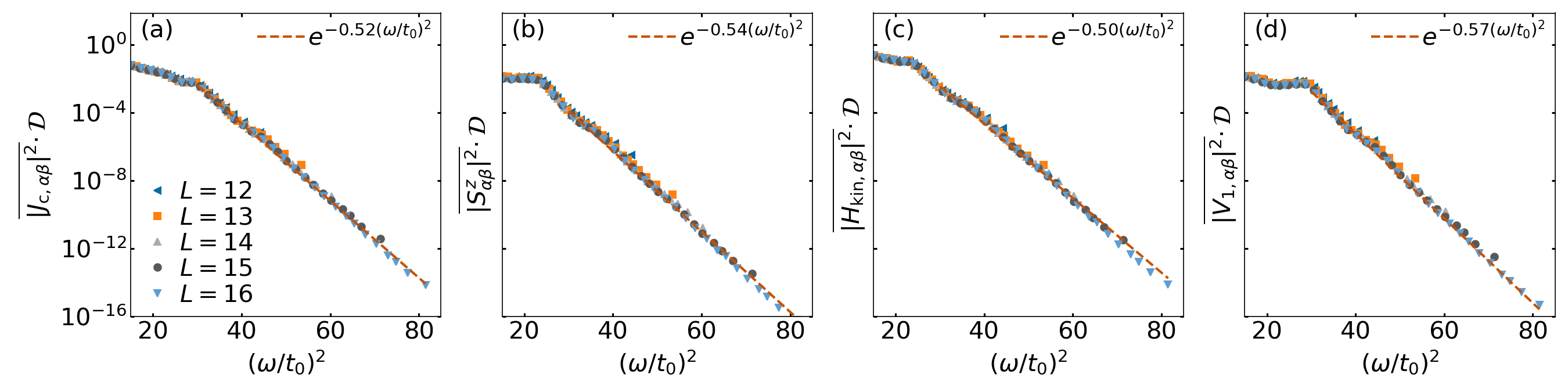}
\caption{
Scaled offdiagonal matrix elements $\overline{|O_{\alpha\beta}|^2} {\cal D}$ as a function of $(\omega/t_0)^2$ for four observables and different system sizes $L=12,...,16$.
Dashed lines are fits of a Gaussian function $\propto e^{- \zeta_O (\omega/t_0)^2}$ to the results for $L=16$ and $(\omega/t_0)^2 > 30$, with the values of $\zeta_O$ given in the legends.
}
\label{fig:offdiag_large_w}
\end{figure*}

The offdiagonal matrix elements $\overline{|O_{\alpha\beta}|^2}$ of the observables $\hat T_1$, $\hat V_1$, $\hat S^z$ and $\hat H_{\rm kin}$ are shown in the main panels of Fig.~\ref{fig:offdiag_1w2}.
A  common feature is the existence of the plateau in the limit $\omega/t_0 \to 0$.
Such a plateau also emerges in the structure of the matrix elements of currents in Fig.~\ref{fig:offdiag_current}.
However, in contrast to the currents,  for the observables in Fig.~\ref{fig:offdiag_1w2}, the width $\Omega_O$ of the plateaus (which is of the order $\Omega_O/t_0 \approx 10^{-2}$) seems to decrease with $L$, as indicated by the dashed arrows in the main panels of Fig.~\ref{fig:offdiag_1w2}.
We will further elaborate on this issue below.
Their second common feature is the signature of an $\propto 1/\omega^2$ decay at moderate $\omega/t_0 \approx 1$, as indicated by the dashed dotted lines in the main panels of Fig.~\ref{fig:offdiag_1w2}.
Properties of matrix elements at very large $\omega/t_0 \gg 1$, which also share some common features, are studied in Fig.~\ref{fig:offdiag_large_w} in Sec.~\ref{sec:large_w}.

We note that other observables introduced in Sec.~\ref{sec:observables}, not presented in Figs.~\ref{fig:offdiag_current} and~\ref{fig:offdiag_1w2}, share similarities with the observables shown in Fig.~\ref{fig:offdiag_1w2} in the sense that the width of the $\omega$-independent plateau in the limit $\omega/t_0 \to 0$ exhibits an $L$ dependence.
However, they do not exhibit signatures of an $\propto 1/\omega^2$ decay, and hence they are not studied in more detail in this section.
In fact, even for the kinetic energy $\hat H_{\rm kin}$ in Fig.~\ref{fig:offdiag_1w2}(d), the $1/\omega^2$ decay appears to be less pronounced.
Still, it exhibits an exponential decay of its autocorrelation function (to be discussed in Sec.~\ref{sec:current_auto}), and hence we study the structure of the offdiagonal matrix elements of $\hat H_{\rm kin}$ along with those for the observables $\hat T_1$, $\hat V_1$, $\hat S^z$.

\begin{figure}[!]
\includegraphics[width=1.0\columnwidth]{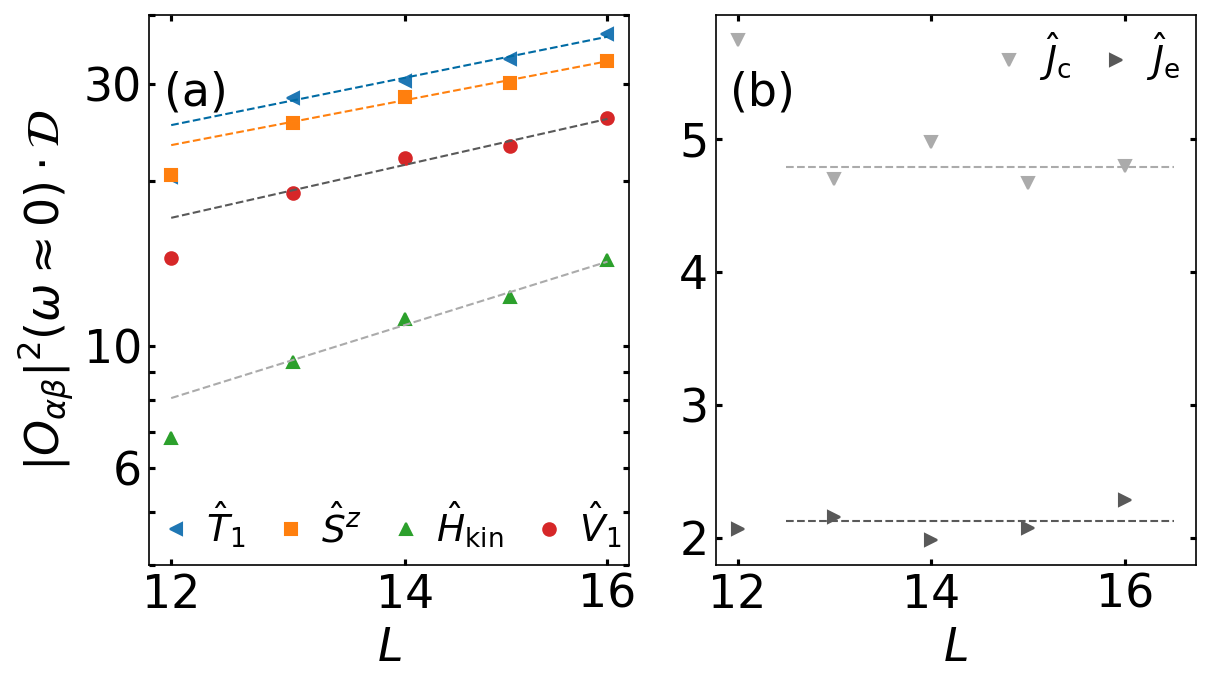}
\caption{
Scaled offdiagonal matrix elements $\overline{|O_{\alpha\beta}|^2} {\cal D}$ at $\omega \approx 0$ as a function of $L$ for (a) the observables $\hat T_1$, $\hat V_1$, $\hat S^z$ and $\hat H_{\rm kin}$ and (b) for the currents $\hat J_{\rm c}$ and $\hat J_{\rm e}$.
The values of $\overline{|O_{\alpha\beta}|^2} {\cal D}$ at $\omega \approx 0$ correspond to the smallest nonzero $\omega$ according to Eq.~(\ref{def_structure_offdiag}), i.e., they are averaged over 700, 1000, 2000, 3000 and 4000 matrix elements for $L=12,13,14,15,16$, respectively.
Dashed lines are fits to the results for $13 \leq L \leq 16$.
In (a), we fit the function $\propto L^\gamma$, where $\gamma = 1.3, 1.4, 1.2, 2.0$ for $\hat T_1$, $\hat V_1$, $\hat S^z$ and $\hat H_{\rm kin}$, respectively.
In (b), we fit a constant (dashed lines) to the data.
}
\label{fig:offdiag_plateau_hight}
\end{figure}

In the insets of Fig.~\ref{fig:offdiag_1w2}, we further illustrate  the scaling properties of the algebraic $\propto 1/\omega^2$ decay of the matrix elements.
We find a good data collapse for different $L$ if the matrix elements are scaled as $\overline{|O_{\alpha\beta}|^2} {\cal D} \omega^2 L$.
Such a scaling suggests that
\begin{equation} \label{def_1omega2}
 \overline{|O_{\alpha\beta}|^2} \, {\cal D} \propto  \frac{1}{L \, \omega^2} \,.
\end{equation}
However, some care is required when applying Eq.~(\ref{def_1omega2}) to the numerical results.
First, it implies that the scaled matrix elements $\overline{|O_{\alpha\beta}|^2} {\cal D} \omega^2 L$ are independent of $\omega$.
The results in the insets of Figs.~\ref{fig:offdiag_1w2}(b) and~\ref{fig:offdiag_1w2}(c) suggest that this may indeed be the case for the observables $\hat V_1$ and $\hat S^z$ for $1 \lesssim \omega/t_0 \lesssim 4$, while the inset of Fig.~\ref{fig:offdiag_1w2}(a) shows a persistent increase of $\overline{|O_{\alpha\beta}|^2} {\cal D} \omega^2 L$ with $\omega$ for $\hat T_1$, eventually suggesting an additional contribution of the form $1/\omega^\zeta$, with $\zeta < 2$.
Finally, the matrix elements of $\hat H_{\rm kin}$ exhibit a peak around $\omega/t_0 \approx 4$ [see the arrow in Fig.~\ref{fig:offdiag_1w2}(d)], which screens a possible $\propto 1/\omega^2$ decay and as a consequence, we do not observe any $\omega$-independent regime when carrying out the rescaling $\overline{|O_{\alpha\beta}|^2} {\cal D} \omega^2 L$ for $\hat H_{\rm kin}$.

Coming back to the properties in the $\omega \to 0$ limit, we note that not only the width of the plateau of the scaled matrix elements $\overline{|O_{\alpha\beta}|^2} {\cal D}$ is $L$ dependent, but also its height.
The latter is at the focus of the analysis presented in Fig.~\ref{fig:offdiag_plateau_hight}, where we plot the scaled matrix elements $\overline{|O_{\alpha\beta}|^2} {\cal D}$ in the $\omega\to 0$ limit as a function of $L$.
In general, those matrix elements increase as $\propto L^\gamma$, where $1\lesssim \gamma \lesssim 2$ for $\hat T_1$, $\hat V_1$, $\hat S^z$ and $\hat H_{\rm kin}$ [see Fig.~\ref{fig:offdiag_plateau_hight}(a)], in contrast to the currents where $\gamma \approx 0$ [see Fig.~\ref{fig:offdiag_plateau_hight}(b)].
In particular, we get $\gamma=1.2$ for $\hat T_1$ and $\gamma=1.3$ for $\hat S^z$, which may suggest ballistic dynamics of those observables ($\gamma \to 1$) in the thermodynamic limit.
However, we obtain $\gamma = 2.0$ for $\hat H_{\rm kin}$, which may be consistent with diffusion.
We also note that the model under investigation is studied in the single-fermion sector, which may impact the scalings of matrix elements in the $L\to \infty$ limit as the density goes to zero.

\begin{figure}[!]
\includegraphics[width=1.0\columnwidth]{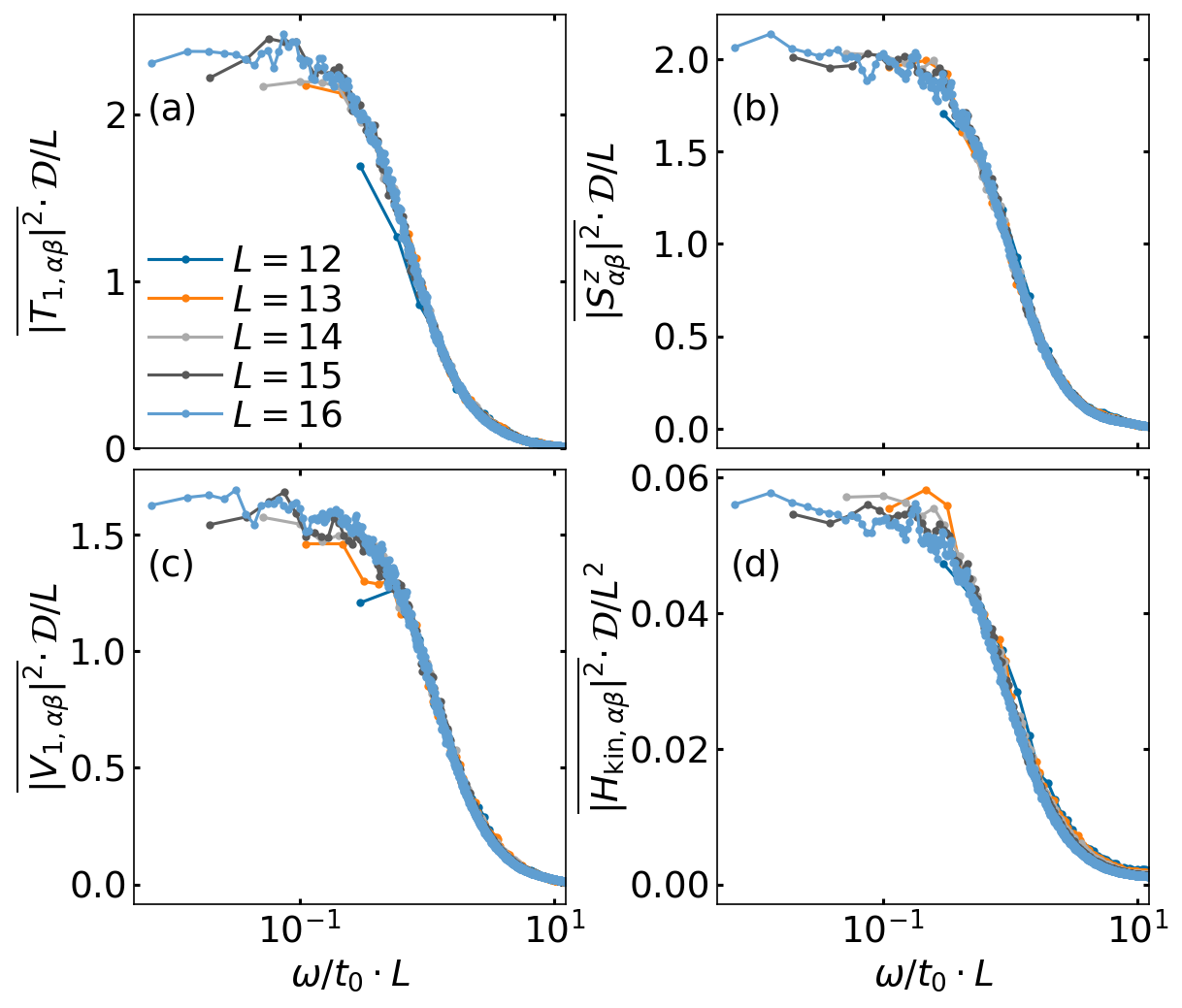}
\caption{
(a)-(c) Scaled offdiagonal matrix elements $\overline{|O_{\alpha\beta}|^2} {\cal D}/L$ for the operators $\hat T_1$, $\hat S^z$ and $\hat V_1$, respectively, and (d) $\overline{|O_{\alpha\beta}|^2} {\cal D}/L^2$ for the fermion kinetic energy $\hat H_{\rm kin}$.
Results are shown as a function of $(\omega/t_0)L$ for different lattice sizes from $L=12$ to $L=16$.
}
\label{fig:offdiag_linear}
\end{figure}

The scaling behaviors studied so far raise the question about a possible phenomenological description of the offdiagonal matrix elements over a wide regime of $\omega/t_0$ that includes the $\omega/t_0 \to 0$ limit and the $\propto 1/(L\omega^2)$ decay at $\omega/t_0 \approx 1$.
A possible ansatz is given by the Drude-like (Lorentzian) function
\begin{equation} \label{def_lorentzian}
 \overline{|O_{\alpha\beta}|^2} \, {\cal D} \propto \frac{\frac{a}{L}}{\left(\frac{a}{L}\right)^2 + \omega^2} \,,
\end{equation}
where $a$ is a constant.
It correctly reproduces the algebraic decay from Eq.~(\ref{def_1omega2}) at $\omega \gg a/L$ and predicts the scaling $\overline{|O_{\alpha\beta}|^2} \, {\cal D} \propto L$ at $\omega/t_0\to 0$, in reasonable agreement with the behavior of the observables $\hat T_1$, $\hat S_z$ and $\hat V_1$ in Fig.~\ref{fig:offdiag_plateau_hight}(a).

If the ansatz from Eq.~(\ref{def_lorentzian}) is applicable, then the scaled matrix elements $\overline{|O_{\alpha\beta}|^2} {\cal D}/L$ plotted versus $(\omega/t_0)L$ should be $L$ independent.
We test that conjecture in Figs.~\ref{fig:offdiag_linear}(a)-\ref{fig:offdiag_linear}(c) for $\hat T_1$, $\hat S_z$ and $\hat V_1$.
Our results indeed show a reasonably good data collapse and hence support the applicability of the Drude-like form in Eq.~(\ref{def_lorentzian}).
Furthermore, we stress that the ansatz in Eq.~(\ref{def_lorentzian}) also describes the $L$ dependence of the plateau width $\Omega_0$ as $\omega \to 0$, since the width of the Lorentzian function scales as $\Omega_O \propto 1/L$.

However, for $\hat H_{\rm kin}$, where $\gamma = 2.0$ in Fig.~\ref{fig:offdiag_plateau_hight}(a),
one needs to multiply the r.h.s.~of Eq.~(\ref{def_lorentzian}) by $L$
to describe the $\overline{|O_{\alpha\beta}|^2} {\cal D} \propto L^2$ dependence in the $\omega \to 0$ limit.
This functional dependence is verified in Fig.~\ref{fig:offdiag_linear}(d) and it yields a rather good data collapse, even though we do not exclude other possibilities for the scaling relation.

To summarize the results of Secs.~\ref{sec:current} and~\ref{sec:drude}, the first remarkable message is that 
the $L$ dependence of the scaled offdiagonal matrix elements $\overline{|O_{\alpha\beta}|^2} {\cal D}$ for the currents is, in the $\omega/t_0 \to 0$ limit, very different from all other observables under investigation.
The second message is that we identified a Drude-like structure of $\overline{|O_{\alpha\beta}|^2} {\cal D}$ for several observables under investigation in the regime of small and moderate values of $\omega/t_0$.

One may wonder what is the common mechanism that gives rise to the Drude-like structure of the offdiagonal matrix elements of the observables studied here.
We note that the three observables ($\hat T_1$, $\hat S^z$ and $\hat H_{\rm kin}$) are parts of the Hamiltonian at the integrable point [$g=0$ in Eq.~(\ref{def_H})], while the other observable ($\hat V_1$) commutes with it.
Using the language of Ref.~\cite{pandey_claeys_20} one can characterize those observables as integrability-preserving observables.
It is then plausible that the departure from integrability by increasing $g$ gives rise to a Drude-like structure of matrix elements~\cite{leblond_sels_21, vidmar_krajewski_21}.
Establishing a systematic correlation between the classes of observables and the functional forms of their offdiagonal matrix elements is beyond the scope of this work.

\subsection{Large-$\omega$ behavior} \label{sec:large_w}

We complement our analysis of the structure of offdiagonal matrix elements with a discussion of their properties at large $\omega/t_0 \gg 1$.
While the observables may exhibit rather distinct features at small and moderate $\omega/t_0$, they appear to exhibit a pretty generic behavior at large $\omega/t_0$.
This is demonstrated for four different observables in Fig.~\ref{fig:offdiag_large_w} (the other observables, not shown here, exhibit a qualitatively similar behavior).
We fit the results using a Gaussian function,
\begin{equation} \label{def_gauss}
 \overline{|O_{\alpha\beta}|^2} \, {\cal D} \propto e^{-\zeta_O (\omega/t_0)^2} \,.
\end{equation}
Figure~\ref{fig:offdiag_large_w} shows the resulting Gaussian functions as dashed lines using the optimal coefficients $\zeta_O$ that are very similar for all observables.
We note, however, that we cannot exclude the existence of a possible $L$-dependent contribution in Eq.~(\ref{def_gauss}), in analogy to the scaling in Eq.~(\ref{def_1omega2}).

The Gaussian decay of the structure of offdiagonal matrix elements appears to be consistent with results of~\cite{jansen_stolpp_19}, which studied the Holstein-polaron model.
Moreover, a recent study of the integrable Heisenberg spin-1/2 chain in the $S^z =0$ total magnetization sector found an exponential decay with $\omega$ at moderate values of $\omega$, followed by a Gaussian decay at large $\omega$~\cite{leblond_mallayya_19}.
In the nonintegrable regime of the Heisenberg spin-1/2 chain, however, only the exponential decay was observed~\cite{leblond_mallayya_19}.
The origin of the difference between the two functional forms needs to be further explored in future work, as well as their relation to other functional forms such as $e^{-\zeta \omega \ln \omega}$ that was recently proposed for the transverse and longitudinal field Ising model~\cite{avdoshkin_dymarsky_20, cao_21}.

\section{Variances of matrix elements}
\label{sec:variances}

We now study fluctuations of both diagonal and offdiagonal matrix elements of observables.
We first focus on the offdiagonal matrix elements.
We define the variance over all offdiagonal matrix elements as
\begin{equation} \label{def_variance_offdiag}
\langle |O_{\alpha\beta}|^2 \rangle_{\cal D} = \frac{1}{{\cal D} ({\cal D}-1)} \sum_{\substack{\alpha',\beta'=1\\ \alpha'\neq\beta'}}^{\cal D} |O_{\alpha'\beta'}|^2 \,,
\end{equation}
where we neglected the contribution from the mean squared, which is of the order $10^{-9}$ or smaller for the system sizes under investigation.

\begin{figure}[b]
\includegraphics[width=0.99\columnwidth]{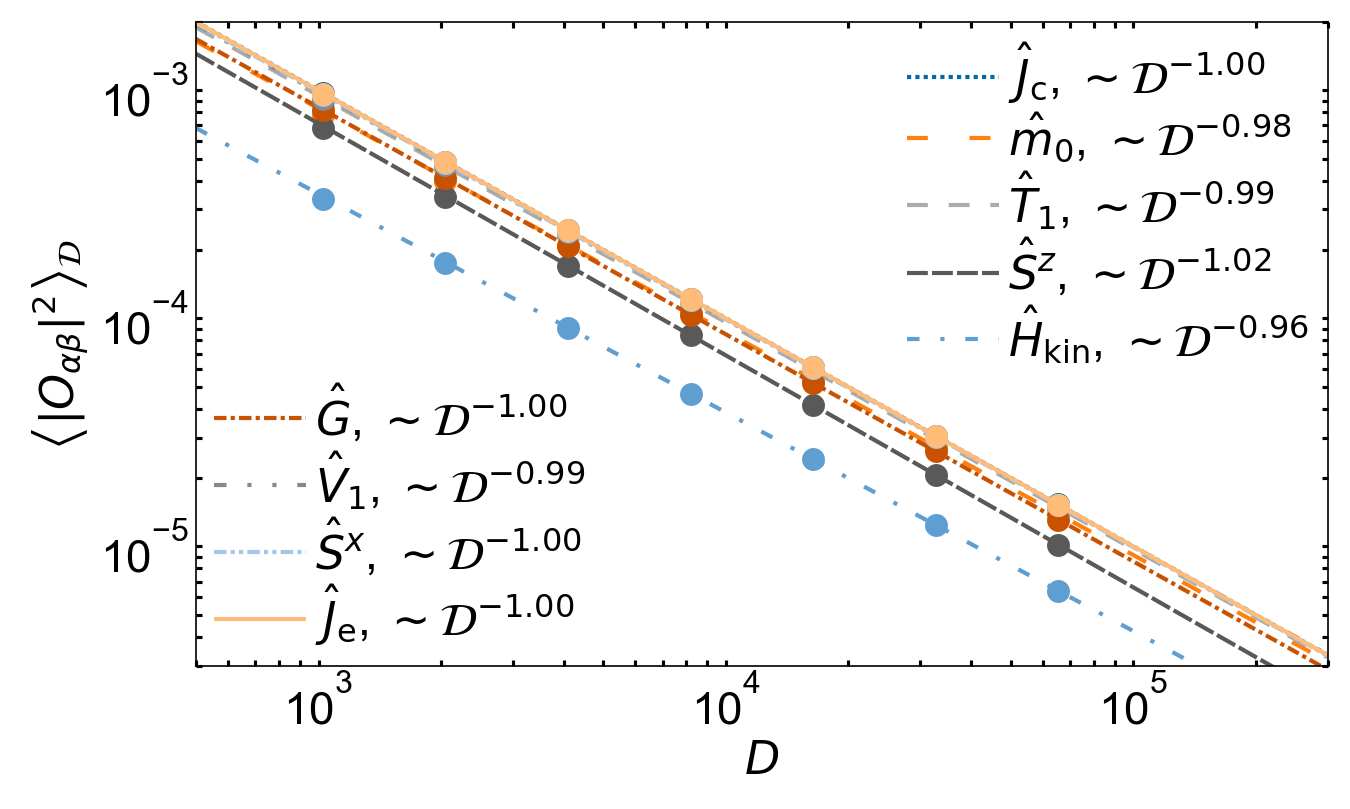}
\caption{
Variances of the offdiagonal matrix elements $\langle |O_{\alpha\beta}|^2 \rangle_{\cal D}$, see Eq.~(\ref{def_variance_offdiag}), for the nine observables defined in Sec.~\ref{sec:observables}.
Symbols are numerical results for systems with $L=10, ..., 16$, while the lines are fits of the function $a_0 {\cal D}^{-\gamma}$ to the data for $L \geq 13$.
The resulting values of $\gamma$ are shown in the figure legend for each observable.
}
\label{fig:var_offdiag}
\end{figure}

The scaling of variances $\langle |O_{\alpha\beta}|^2 \rangle_{\cal D}$ versus the Hilbert-space dimension ${\cal D}$ is shown as symbols in Fig.~\ref{fig:var_offdiag}, while lines represent fits of the data to the function $a_0 {\cal D}^{-\gamma}$.
We observe $\gamma \approx 1$ for all nine observables defined in Sec.~\ref{sec:observables}, in accord with the ETH ansatz in Eq.~(\ref{def_eth_ansatz}).
The numerical accuracy of $\gamma$ is on the second digit.
This represents a remarkable manifestation of the ETH in finite systems, which can be studied via exact diagonalization.

The results of Fig.~\ref{fig:var_offdiag} for the spin-fermion model are consistent with previous studies of variances for Heisenberg-like spin Hamiltonians~\cite{leblond_mallayya_19}.
We note that here, we use the same normalization of observables as in~\cite{mierzejewski_vidmar_20}, for which the ETH ansatz is given by Eq.~(\ref{def_eth_ansatz}).
For intensive observables of the form $\hat O = (1/L) \sum_j \hat o_j$, where $\hat o_j$ is a local operator around site $j$ with support on ${\cal O}(1)$ sites, the Hilbert-Schmidt norm of observables~(\ref{def_norm}) typically decays as $~1/L$.
It has been argued in~\cite{leblond_mallayya_19} that, for intensive observables, the second part of the ETH ansatz~(\ref{def_eth_ansatz}) needs to be multiplied by $1/\sqrt{L}$ to yield the same scaling of the variances as observed in Fig.~\ref{fig:var_offdiag}.

Interestingly, the variances in Fig.~\ref{fig:var_offdiag} do not only exhibit a nearly identical exponent $\gamma$ in the fitting function $a_0 {\cal D}^{-\gamma}$, but also the prefactor $a_0$ may be very similar.
In fact, we observe $a_0 = \gamma = 1.00$ for the charge and energy currents $\hat J_{\rm c}$ and $\hat J_{\rm e}$.
The latter result implies that
\begin{equation}
 \frac{1}{\cal D} \sum_{\substack{\alpha,\beta=1\\ \alpha\neq\beta}}^{\cal D} |O_{\alpha\beta}|^2 \approx 1 \,,
\end{equation}
with exponentially small corrections.
This is a consequence of the observable normalization introduced in Sec.~\ref{sec:observables} and a vanishing contribution of the diagonal matrix elements to the operator norm,
$1/{\cal D} \sum_\alpha O_{\alpha\alpha}^2 \ll 1$.
This is the case for observables whose diagonal matrix elements are structureless, such as $\hat J_{\rm c}$ and $\hat J_{\rm e}$ shown in Figs.~\ref{fig:eev1}(a) and~\ref{fig:eev1}(b), respectively.

Note that the prefactor $a_0$ is noticeably different (specifically, $a_0 < 1$) for the observables $\hat H_{\rm kin}$, $\hat S^z$ and $\hat G$, which are all part of the Hamiltonian~(\ref{def_H}).
This can be understood along the lines of Ref.~\cite{mierzejewski_vidmar_20}:
if an observable is part of the Hamiltonian,  the contribution of the diagonal matrix elements to the observable normalization~(\ref{def_norm}) is considerable (since the projection of the observable on the Hamiltonian is large, as discussed in Sec.~\ref{sec:diagonals}).
Therefore, to fulfill the observable normalization, the contribution of the offdiagonal matrix elements needs to decrease, which results in a lower value of $a_0$.

We contrast the variances of the offdiagonal matrix elements with those of the diagonal ones.
For the latter, however, several definitions of variances were used in the past.
For the diagonal matrix elements of observables with nonvanishing structure, that structure needs to be subtracted in an appropriate microcanonical window (see, e.g., Refs.~\cite{beugeling_moessner_14, yoshizawa_iyoda_18, mierzejewski_vidmar_20}).
Another possible measure of the fluctuations that does not require any subtraction are eigenstate-to-eigenstate fluctuations of matrix elements~\cite{kim_ikeda_14}, which turned out to be a powerful indicator of the validity of the ETH ansatz~\cite{mondaini_fratus_16, jansen_stolpp_19, leblond_mallayya_19}.
However, here we focus on observables whose diagonal matrix elements have no structure [cf.~$\hat J_{\rm c}$, $\hat J_{\rm e}$ and $\hat V_1$, see Figs.~\ref{fig:eev1}(a)-\ref{fig:eev1}(c)], and hence the subtraction of the mean values in microcanonical windows is not necessary~\cite{mierzejewski_vidmar_20}.
This allows us to use a simple definition of the variance of diagonal matrix elements,
\begin{equation} \label{def_variance_diag}
\langle O_{\alpha\alpha}^2 \rangle_\mu = \frac{1}{\mu} \sum_{\alpha' = {\cal D}/2 - \mu/2 + 1}^{{\cal D}/2 + \mu/2} (O_{\alpha' \alpha'})^2 \,,
\end{equation}
where the role of $\mu$ is to remove contributions from spectral edges.
The dominant contribution to the variance is expected to decay as $\propto 1/{\cal D}$ if the system satisfies ETH, and hence we study the scaled variances $\langle O_{\alpha\alpha}^2 \rangle_\mu {\cal D}$ further on.

The convenience of studying structureless observables can also be understood by the analysis of how the expectation values of observables in the diagonal ensemble~\cite{rigol_dunjko_08} approach the corresponding expectation value in the microcanonical ensemble~\cite{ikeda_ueda_15, dalessio_kafri_16}.
In this analysis, one typically expands the structure function $O(E)$ in a power series around the target energy $E$, which gives rise to the differences between the diagonal and the microcanonical ensemble averages that vanish polynomially with $L$.
In contrast, if $O(E) = 0$, those polynomial contributions are zero, giving rise to the leading term that scales algebraically with ${\cal D}$ (i.e., it is exponentially small in $L$).

\begin{figure}[!]
\includegraphics[width=0.99\columnwidth]{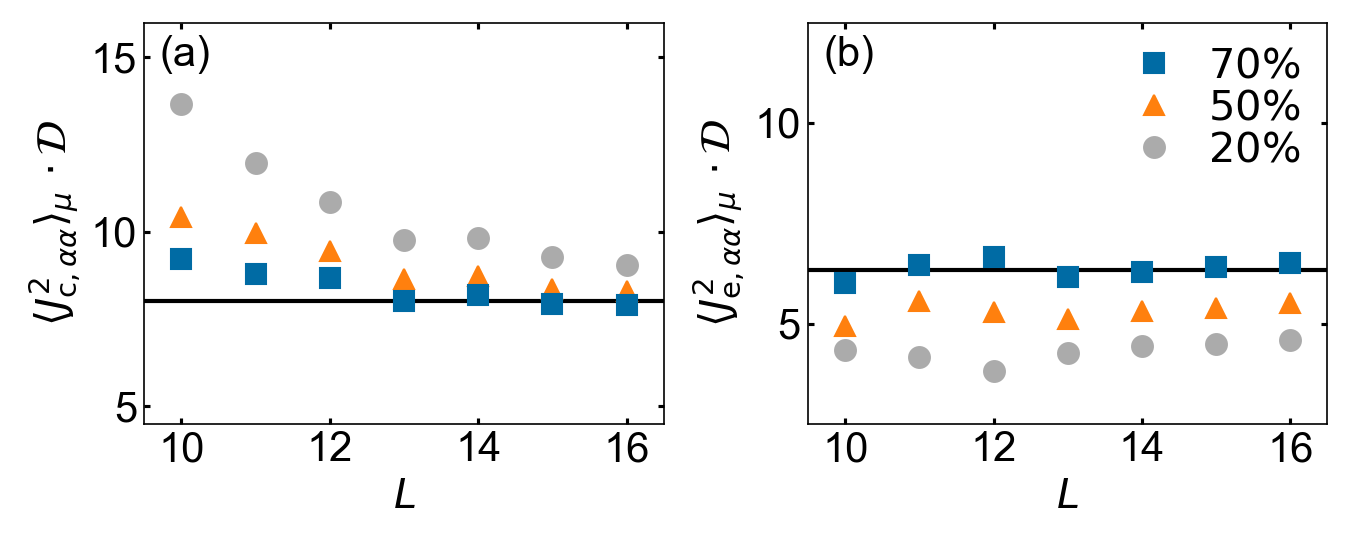}
\caption{
Scaled variances of the diagonal matrix elements $\langle O_{\alpha\alpha}^2 \rangle_\mu \, {\cal D}$ for (a) the charge current $\hat J_{\rm c}$ and (b) the energy current $\hat J_{\rm e}$, as a function of the lattice size $L$.
Lines are fits of the data for $L \geq 13$ to a constant.
We choose $\mu$ in Eq.~(\ref{def_variance_diag}) such that we include results for 70\%, 50\% and 20\% eigenstates around the center of the spectrum (see the legend).
}
\label{fig:var_diag_J}
\end{figure}

\begin{figure}[!]
\includegraphics[width=0.99\columnwidth]{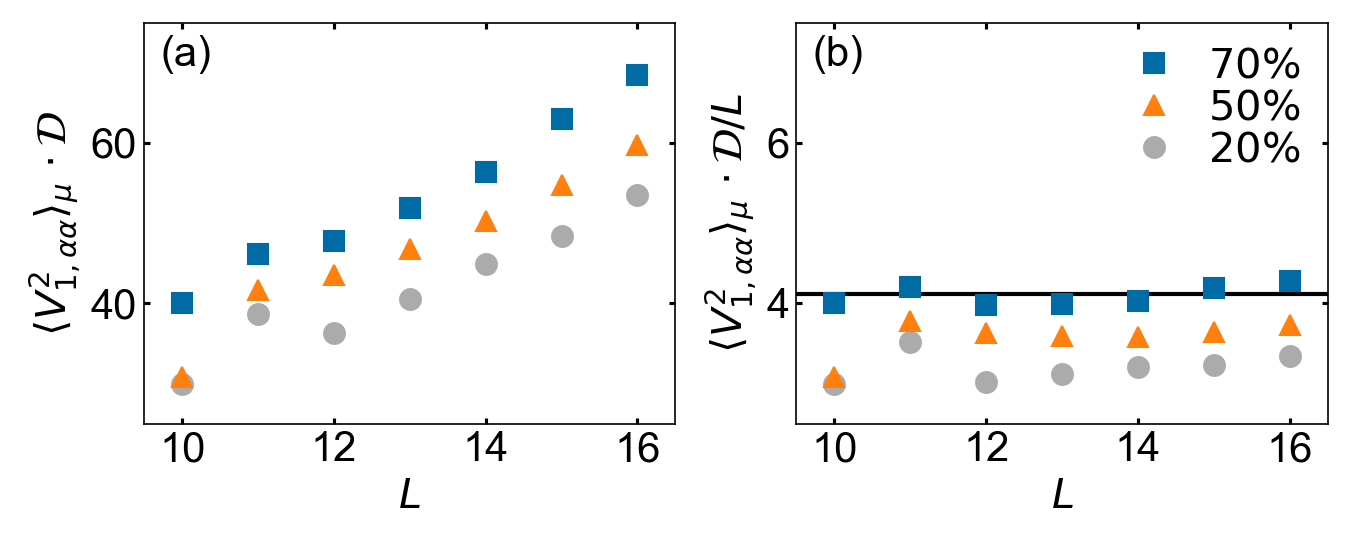}
\caption{
Scaled variances of the diagonal matrix elements for the spin correlator $\hat V_1$ as a function of the lattice size $L$.
Variances are scaled as (a) $\langle V_{1,{\alpha\alpha}}^2 \rangle_\mu \, {\cal D}$ and (b) $\langle V_{1,{\alpha\alpha}}^2 \rangle_\mu \, {\cal D}/L$.
The line in (b) is a fit to a constant for $L \geq 13$.
We choose $\mu$ in Eq.~(\ref{def_variance_diag}) such that we include results for 70\%, 50\% and 20\% eigenstates around the center of the spectrum (see legend).
}
\label{fig:var_diag_V1}
\end{figure}

\begin{figure*}[!]
\includegraphics[width=1.5\columnwidth]{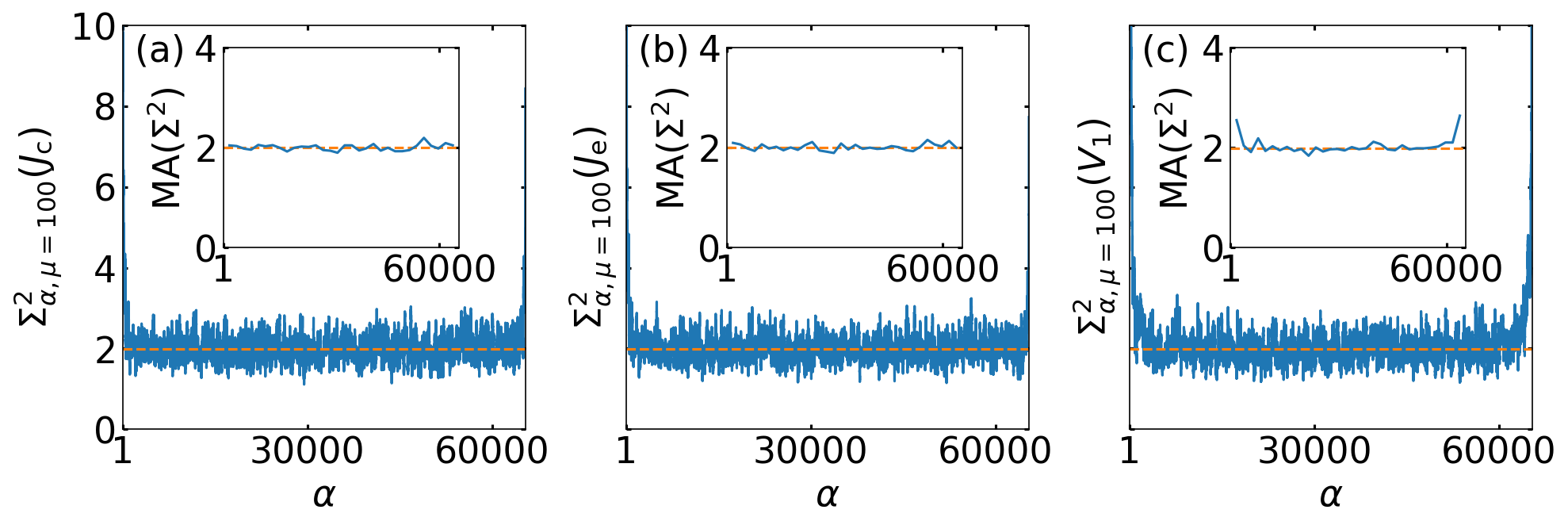}
\caption{
Ratio of variances of diagonal versus offdiagonal matrix elements $\Sigma_{\alpha,\mu=100}^2(O)$, see Eq.~(\ref{def_variance_ratio}), at $L=16$ as a function of the eigenstate index $\alpha$.
Each panel includes $2^{16} - 99$ values of $\Sigma_{\alpha,\mu=100}^2(O)$.
In the inset, the moving average (MA) is performed over the 2000 nearest values.
Results are shown for the operators $\hat J_{\rm c}$, $\hat J_{\rm e}$, and $\hat V_1$ in (a)-(c), respectively.
Horizontal dashed lines represent the averages $\overline{\Sigma_{\mu=100}^2}(O)$, which are obtained by averaging over those sets of ratios of variances for which the mean energy $\overline{E_\alpha} = \sum_{\rho = \alpha}^{\alpha+99} E_\rho/100$ lies in the window $\overline{E_\alpha}/(L t_0) \in [-1/6,1/6]$
(such average includes 83.6 $\%$ of all values in the main panel).
We get $\overline{\Sigma_{\mu=100}^2}(J_{\rm c}) = 1.994$, $\overline{\Sigma_{\mu=100}^2}(J_{\rm e}) = 1.995$ and $\overline{\Sigma_{\mu=100}^2}(V_1) = 1.984$.
}
\label{fig:var_ratio}
\end{figure*}

Figure~\ref{fig:var_diag_J} shows the scaled variances $\langle O_{\alpha\alpha}^2 \rangle_\mu {\cal D}$ of the currents $\hat J_{\rm c}$ and $\hat J_{\rm e}$, as functions of the lattice size $L$.
They appear to be nearly $L$ independent.
This is similar to the scaling of fluctuations of their offdiagonal matrix elements in a narrow energy window at $\omega/t_0 \to 0$, see Fig.~\ref{fig:offdiag_plateau_hight}(b).
In fact, we argue that the results in Fig.~\ref{fig:offdiag_plateau_hight}(b) and~\ref{fig:var_diag_J} are consistent with each other since the ratio of variances [to be defined in Eq.~(\ref{def_variance_ratio})] is an $L$-independent value $\Sigma^2=2$, as shown in Figs.~\ref{fig:var_ratio}(a)-\ref{fig:var_ratio}(b).

In contrast to the currents, the scaled variance of the diagonal matrix elements of the spin correlator $\hat V_1$, see Fig.~\ref{fig:var_diag_V1}(a), keeps increasing as a function of $L$ for the system sizes under investigation.
One possible interpretation of the results in Fig.~\ref{fig:var_diag_V1}(a) is, along the lines of Ref.~\cite{yoshizawa_iyoda_18}, to argue that the variance scales as $\langle V_{1,\alpha\alpha}^2 \rangle_\mu \propto {\cal D}^{-z}$, with $z<1$.
However, here we interpret the numerical results as a polynomial increase of $\langle V_{1,\alpha\alpha}^2 \rangle_\mu {\cal D}$ with $L$.
This is corroborated in Fig.~\ref{fig:var_diag_V1}(b), where we show that the scaled variance $\langle V_{1,\alpha\alpha}^2 \rangle_\mu {\cal D}/L$ appears to be approximately $L$ independent.
While the extraction of the precise functional form of the $L$ dependence cannot be determined from the available system sizes, we argue that the scaling of the diagonals in Fig.~\ref{fig:var_diag_V1}(a) is in agreement with the scaling of the variance of the offdiagonal matrix elements at $\omega/t_0 \to 0$ shown in Fig.~\ref{fig:offdiag_plateau_hight}(a), where it was observed that $\overline{|V_{1,\alpha\beta}|^2}{\cal D} \propto L^{\gamma}$, with $\gamma=1.4$.

The results presented so far can be summarized as follows:
(i) the system-size dependence of the variances of the diagonal matrix elements may strongly depend on the observable,
and (ii) the variances of the diagonal matrix elements seem to exhibit a very similar scaling as the variances of the corresponding offdiagonal matrix elements in the $\omega/t_0 \to 0$ limit.
The latter observation is further corroborated by studying the ratio of variances, defined as
\begin{equation} \label{def_variance_ratio}
 \Sigma_{\alpha,\mu}^2(O) = \frac{\left[ \sigma_{\rm diag}^{(\alpha,\mu)}(O) \right]^2}{\left[ \sigma_{\rm offdiag}^{(\alpha,\mu)}(O) \right]^2} \,,
\end{equation}
where the variances of the diagonal and offdiagonal matrix elements, 
\begin{equation}
 \left[ \sigma_{\rm diag}^{(\alpha,\mu)}(O) \right]^2 = \frac{1}{\mu} \sum_{\rho =\alpha}^{\alpha+\mu-1} O_{\rho\rho}^2 - \left( \frac{1}{\mu}\sum_{\rho =\alpha}^{\alpha+\mu-1} O_{\rho\rho}\right)^2 \label{def_variance_diag_ratio}
\end{equation}
and
\begin{equation}
 \left[ \sigma_{\rm offdiag}^{(\alpha,\mu)}(O) \right]^2 = \frac{1}{\mu^2 - \mu} \sum_{\substack{\rho,\rho' = \alpha \\ \rho \neq \rho'}}^{\alpha+\mu-1} |O_{\rho\rho'}|^2 \,,
\end{equation}
respectively, are now defined for a set of $\mu$ consecutive eigenstates starting at the eigenstate $|\alpha\rangle$.

Based on the random-matrix theory, it was argued that $\Sigma^2 = 2$ in many-body systems that comply with the ETH~\cite{dalessio_kafri_16}.
Indeed, results of large-scale numerical calculations for matrix elements of observables in the transverse field Ising model~\cite{mondaini_rigol_17} and the Holstein-polaron model~\cite{jansen_stolpp_19} are consistent with this expectation.
Those studies also highlight that the expected result can only be observed for small $\mu$, i.e., in the limit $\omega/t_0 \to 0$ of the offdiagonal matrix elements.
The latter statement is independent of whether the structure of the diagonal matrix elements of observables is vanishingly small or not.

In Fig.~\ref{fig:var_ratio}, we show the ratio $\Sigma_{\alpha,\mu}^2(O)$ for the observables $\hat J_{\rm c}$, $\hat J_{\rm e}$ and $\hat V_1$.
Note that for these observables, the subtraction of the square of the mean value on the r.h.s. in Eq.~(\ref{def_variance_diag_ratio}) is used only to improve the numerical accuracy and can be omitted.
The results show that the ratio of variances is indeed very close to 2 in finite systems.
Hence, in the limit $\omega \to 0$, the variance of the offdiagonal matrix elements should have an identical system-size scaling as the variance of the diagonal matrix elements.

It is remarkable how good the agreement between $\Sigma_{\alpha,\mu}^2(O)$ and the random matrix theory prediction is.
For the currents studied in Figs.~\ref{fig:var_ratio}(a) and~\ref{fig:var_ratio}(b), we get the numerical values
$\overline{\Sigma_{\mu=100}^2}(J_{\rm c}) = 1.994$ and $\overline{\Sigma_{\mu=100}^2}(J_{\rm e}) = 1.995$, where the average is performed over those sets of ratios of variances for which the mean energy $\overline{E_\alpha} = \sum_{\rho = \alpha}^{\alpha+99} E_\rho/100$ lies in the window $\overline{E_\alpha}/(L t_0) \in [-1/6,1/6]$, see Fig.~\ref{fig:var_ratio} for details.

Summarizing the analysis of the matrix-element fluctuations, we argue that the most generic feature is the result for the ratio of their variances, $\Sigma_{\alpha,\mu}^2(O) \approx 2$ in narrow energy windows ($\omega/t_0 \to 0$ for the offdiagonals).
However, the ETH ansatz {\it per se} does not imply any particular scaling of the fluctuations with system size $L$ (beyond the dominant exponential suppression), which is illustrated by the difference between the matrix elements of currents and other observables.

\section{Spectral densities of operators}
\label{sec:spectrum}

Next, we study properties of the spectral densities of observables, which are related to the autocorrelation functions, such as the one in Eq.~(\ref{def_C_alpha}), by a Fourier transform.
For the symmetric autocorrelation function $C_{O}^{(\alpha)}(t)$ defined in Eq.~(\ref{def_C_alpha}), the spectral density $S_{O,+}^{(\alpha)}(\omega)$ is defined in Eq.~(\ref{def_S_alpha}).
It consists of two contributions,
\begin{equation} \label{def_S_alpha_plus}
 S_{O,+}^{(\alpha)}(\omega) = S_{O}^{(\alpha)}(\omega) + S_{O}^{(\alpha)}(-\omega) \,,
\end{equation}
where
\begin{align} \label{def_S_alpha_positive}
 S_{O}^{(\alpha)}(\omega) & = \int_{-\infty}^\infty dt \, e^{i \omega t} \, \langle \alpha |\hat O(t) \hat O(0) |\alpha \rangle_{\rm c} \,, \\
 S_{O}^{(\alpha)}(-\omega) & = \int_{-\infty}^\infty dt \, e^{i \omega t} \, \langle \alpha |\hat O(0) \hat O(t) |\alpha \rangle_{\rm c} \,. \label{def_S_alpha_omeganeg}
\end{align}
In Eq.~(\ref{def_S_alpha_omeganeg}), invariance under time translations has been assumed.
Writing the operators  in Eq.~(\ref{def_S_alpha_positive}) in the Heisenberg picture, inserting a complete eigenbasis $\hat I = \sum_\beta |\beta\rangle\langle\beta|$, and performing the integral over time one can express $S_{O}^{(\alpha)}(\omega)$ by the matrix elements of observables as
\begin{align}
 S_{O}^{(\alpha)}(\omega) & = 2\pi \sum_{\beta \neq \alpha} |O_{\alpha\beta}|^2 \delta[\omega-(E_\beta - E_\alpha)] \nonumber \\
 & = 2\pi \; \overline{|O_{E_\alpha,E_{\alpha}+\omega}|^2} \; \rho(E_\alpha+\omega) \,.
 \label{def_S1_alpha}
\end{align}
In the derivation of the final result in Eq.~(\ref{def_S1_alpha}),
we averaged the matrix elements over eigenstates $\beta$ such that $|\omega - (E_\beta - E_\alpha)| \leq \delta \omega$, where $\delta \omega$ should be much larger than the mean level spacing (such that it includes $\delta N \gg 1$ states) and much smaller than any other relevant energy scale in the system.
This is consistent with the averaging of the offdiagonal matrix elements used in Eq.~(\ref{def_structure_offdiag}).
The density of states $\rho(E)$ in Eq.~(\ref{def_S1_alpha}) is then defined as $\rho(E) = \delta N/\delta \omega$.

An important consequence of the choice of normalization of observables invoked in Eq.~(\ref{def_norm}) is the nonvanishing value of the sum rule of the spectral density.
For a typical eigenstate $|\alpha\rangle$ in the bulk of the spectrum, one can express it as
\begin{equation} \label{def_sumrule}
 \int_{-\infty}^\infty S_{O,+}^{(\alpha)}(\omega) d\omega = 4\pi \sum_{\beta\neq\alpha} |O_{\alpha\beta}|^2 \approx 4\pi + {\cal O}({\cal D}^{-1}) \,,
\end{equation}
where it is assumed that the contribution of the missing diagonal matrix element in the sum of Eq.~(\ref{def_sumrule}) is ${\cal O}({\cal D}^{-1})$.

\subsection{Fluctuation-dissipation theorem (FDT)}

The spectral density provides access to squares of matrix elements of observables, averaged over a narrow energy window.
In Eqs.~(\ref{def_S_alpha}) and~(\ref{def_S_alpha_plus}), we defined the symmetric eigenstate spectral density $S_{O,+}^{(\alpha)}(\omega)$ through the anticommutator of the two-time correlation function.
One can analogously define the antisymmetric eigenstate spectral density $S_{O,-}^{(\alpha)}(\omega)$ by replacing the anticommutator with the commutator, 
\begin{equation} \label{def_S_alpha_minus}
S_{O,-}^{(\alpha)}(\omega) = S_O^{(\alpha)}(\omega) - S_O^{(\alpha)}(-\omega) \,.
\end{equation}
The latter is relevant for the definition of the response functions.
In particular, the dissipative part of the eigenstate Kubo linear response function can be expressed by $S_{O,-}^{(\alpha)}(\omega)$ as
\begin{equation} \label{def_kubo}
 {\rm Im}\left(\chi_O^{(\alpha)}(\omega)\right) = \int_0^\infty dt e^{i\omega t} \langle \alpha | [\hat O(t), \hat O(0)] | \alpha\rangle_{\rm c} = \frac{S_{O,-}^{(\alpha)}(\omega)}{2} \,.
\end{equation}

The relationship between $S_{O,+}^{(\alpha)}(\omega)$ and $S_{O,-}^{(\alpha)}(\omega)$ represents the core of the fluctuation-dissipation theorem (FDT).
If the spectral densities are evaluated in a Gibbs ensemble, it is well known that the FDT can be derived without assuming any particular form of the matrix elements of observables (see, e.g.,~Ref.~\cite{schwabl_08}).
Remarkably, for systems in which the ETH ansatz~(\ref{def_eth_ansatz}) is valid, the FDT can be derived for a single eigenstate~\cite{dalessio_kafri_16}.
Below we sketch this derivation (see also~\cite{dalessio_kafri_16}).

It is convenient to express the density of states $\rho(E_\alpha \pm \omega)$ through the thermodynamic entropy $S(E_\alpha \pm \omega)$ at the same energy.
Since the eigenstate energy $E_\alpha$ scales extensively with the lattice size $L$ and the target range of $\omega$ does not scale with $L$, one can expand $S(E_\alpha \pm \omega)$ around $S(E_\alpha)$ as
\begin{equation} \label{expand_rho}
 \rho(E_\alpha \pm \omega) = e^{S(E_\alpha\pm \omega)} = e^{S(E_\alpha) \pm \beta \omega + \frac{\omega^2}{2}\frac{\partial \beta}{\partial E_\alpha} + ... } \, ,
\end{equation}
where we define the inverse temperature $\beta$ of an eigenstate $|\alpha\rangle$ as $\beta=\partial S/\partial E_\alpha$, and the density of states is expressed in dimensionless units.

Next, we introduce the averages $\overline{|O_{E_\alpha,E_\alpha\pm \omega}|^2}$ of offdiagonal matrix elements of observables over a narrow window around the target eigenstates at $E_\alpha$ and $E_\beta = E_\alpha \pm \omega$.
We express it using the ETH ansatz~(\ref{def_eth_ansatz}) as
\begin{align} \label{Oab_average}
 \overline{|O_{E_\alpha,E_\alpha\pm \omega}|^2} = &  \frac{|f_O(E_\alpha \pm \frac{\omega}{2},\omega)|^2}{\rho(E_\alpha \pm \omega/2)} \overline{|R_{E_\alpha, E_\alpha\pm \omega}|^2} \,.
\end{align}
The averaging window is assumed to be large enough to set the fluctuating part $\overline{|R_{E_\alpha, E_\alpha\pm \omega}|^2} = 1$, but also narrow enough such that the smooth functions $|f_O|^2$ and $\rho$ are accurately described by the values at the target eigenstates $E_\alpha$ and $E_\beta = E_\alpha \pm \omega$.
We assume the operators to be Hermitian and to satisfy $f(E,\omega)^* = f(E,-\omega)$, which yields $|f(E_\alpha - \frac{\omega}{2},-\omega)|^2 = |f(E_\alpha - \frac{\omega}{2},\omega)|^2$ in Eq.~(\ref{Oab_average}).
Then we expand the remaining quantities around $E_\alpha$,
\begin{align}
 \overline{|O_{E_\alpha,E_\alpha\pm \omega}|^2} = & \left( |f_O(E_\alpha,\omega)|^2 \pm \frac{\omega}{2} \frac{\partial |f_O(E_\alpha,\omega)|^2}{\partial E_\alpha} \right) \nonumber \\
& \times e^{-\left( S(E_\alpha) \pm \frac{\beta\omega}{2} + \frac{\omega^2}{8} \frac{\partial \beta}{\partial E_\alpha}  + ... \right)} \,.
\label{expand_O}
\end{align}
The density of states in Eq.~(\ref{expand_rho}) and the averaged offdiagonal matrix elements in Eq.~(\ref{expand_O}) can now be plugged into Eq.~(\ref{def_S1_alpha}) to obtain
\begin{align}
 S_O^{(\alpha)}(\pm\omega) = \, & 2\pi \left( |f_O(E_\alpha,\omega)|^2 \pm \frac{\omega}{2} \frac{\partial |f_O(E_\alpha,\omega)|^2}{\partial E_\alpha} \right) \nonumber \\ 
 & \times \, e^{\pm\frac{\beta \omega}{2} + \frac{3\omega^2}{8}\frac{\partial \beta}{\partial E_\alpha} + ...} \; . \label{def_S_alpha_expand}
\end{align}
In the latter equation, all quantities are expanded up to the terms that are derivatives with respect to the extensive energy $E_\alpha$ (they are expected to scale as $\propto 1/L$).
The standard form of the fluctuation-dissipation relation is derived by neglecting all such terms.
We follow this approach here, while in Sec.~\ref{sec:fdt_fss}, we also explore finite-size corrections.

The symmetric spectral density from Eq.~(\ref{def_S_alpha_plus}) can be expressed using Eq.~(\ref{def_S_alpha_expand}) as
\begin{equation}
 S_{O,+}^{(\alpha)}(\omega) = 2\pi |f_O(E_\alpha,\omega)|^2 e^{\beta \omega/2} + 2\pi |f_O(E_\alpha,\omega)|^2 e^{-\beta \omega/2} \,,
\end{equation}
resulting is Eq.~(\ref{def_S_alpha_short}), which is, for convenience, repeated below:
\begin{equation} \label{def_S_alpha_short2}
 S_{O,+}^{(\alpha)}(\omega) = 4\pi \cosh\left(\frac{\beta \omega}{2}\right) |f_O(E_\alpha, \omega)|^2 \,.
\end{equation}
Analogously, the antisymmetric spectral density from Eq.~(\ref{def_S_alpha_minus}) can be expressed as
\begin{equation} \label{def_S_alpha_minus_short}
 S_{O,-}^{(\alpha)}(\omega) = 4\pi \sinh\left(\frac{\beta \omega}{2}\right) |f_O(E_\alpha,\omega)|^2 \,.
\end{equation}

Using Eq.~(\ref{def_S_alpha_expand}), one can also obtain another convenient property of the spectral densities, also known as the Kubo-Martin-Schwinger relation~\cite{kubo_57, martin_schwinger_59, haag_hugenholtz_67},
\begin{equation} \label{def_kms}
 S_{O}^{(\alpha)}(\omega) = e^{\beta \omega} \; S_{O}^{(\alpha)}(-\omega) \,.
\end{equation}
In fact, the latter expression is a key and sufficient ingredient for the derivation of the FDT.

Finally, relating the dissipative contributions in Eqs.~(\ref{def_kubo}) and~(\ref{def_S_alpha_minus_short}) to the fluctuations in Eq.~(\ref{def_S_alpha_short2}), one arrives at the fluctuation-dissipation relation
\begin{equation}
 S_{O,+}^{(\alpha)}(\omega) = \coth \left( \frac{\beta \omega}{2} \right) \, 2 \, {\rm Im}\left(\chi_O^{(\alpha)}(\omega)\right)
\end{equation}
as expressed in~\cite{dalessio_kafri_16}.
Another way to express the fluctuation-dissipation relation is
\begin{equation}
 \frac{ S_{O,-}^{(\alpha)}(\omega) }{ S_{O,+}^{(\alpha)}(\omega) } = \tanh \left( \frac{\beta \omega}{2} \right) \,.
\end{equation}
Calculating the time-evolving spectral densities of currents after quantum quenches in the Holstein-polaron model~\cite{kogoj_vidmar_16}, the latter relation was observed after long times, thereby demonstrating the restoration of the fluctuation-dissipation relation in isolated nonequilibrium quantum states.

\subsection{Numerical verification of the FDT}
\label{sec:fdt_numerics}

We complement the previous analytical considerations by numerically verifying the fluctuation-dissipation relation for Hamiltonian eigenstates.
The relation that we actually put to a test is the Kubo-Martin-Schwinger relation from Eq.~(\ref{def_kms}).
We focus on eigenstates in the center of the spectrum, for which $\beta = 0$.

If the FDT is fulfilled for every eigenstate, it implies validity of the Kubo-Martin-Schwinger relation $S_{O}^{(\alpha)}(\omega)/S_{O}^{(\alpha)}(-\omega) = 1$ for each $\alpha$.
Here, we calculate the typical value of the ratio $S_{O}^{(\alpha)}(\omega)/S_{O}^{(\alpha)}(-\omega)$, which is the geometric mean defined as
\begin{equation} \label{def_R_typical}
\log\left( R_{O,\mu}^{(\rm typ)}(\omega) \right) = \frac{1}{\mu} \sum_{\alpha = {\cal D}/2-\mu/2+1}^{{\cal D}/2+\mu/2} \log\left( \frac{S_{O}^{(\alpha)}(\omega)}{S_{O}^{(\alpha)}(-\omega)} \right) \,,
\end{equation}
and the average value of the ratio, which is the arithmetic mean analogous to the one in Eq.~(\ref{def_variance_diag}),
\begin{equation} \label{def_R_average}
R_{O,\mu}^{(\rm avr)}(\omega) = \frac{1}{\mu} \sum_{\alpha = {\cal D}/2-\mu/2+1}^{{\cal D}/2+\mu/2} \frac{S_{O}^{(\alpha)}(\omega)}{S_{O}^{(\alpha)}(-\omega)} \,.
\end{equation}

Note that the calculation of $S_{O}^{(\alpha)}(\pm\omega)$ in Eqs.~(\ref{def_R_typical}) and~(\ref{def_R_average}), using Eq.~(\ref{def_S1_alpha}), requires an averaging over eigenstates $\beta$ whose energies match the condition $|\omega - (E_\beta - E_\alpha)| \leq \delta \omega$.
Hence one needs to average over sufficiently many offdiagonal matrix elements in a window $\delta\omega$ to smoothen fluctuations, which requires $\delta \omega$ to be sufficiently large.
However, $\delta \omega$ should also be sufficiently small to assure high resolution in $\omega$ of the quantities under investigation.
If, for a chosen $\delta \omega$, there is no eigenstate $\beta$ to fulfill this condition, we exclude the eigenstate $\alpha$ from the means in Eqs.~(\ref{def_R_typical}) and~(\ref{def_R_average}).
In this work we partition the $\omega$-axis using equidistant intervals on a log scale, and choose $\delta \omega$ accordingly such that the intervals do not overlap.

\begin{figure}[!]
\includegraphics[width=0.99\columnwidth]{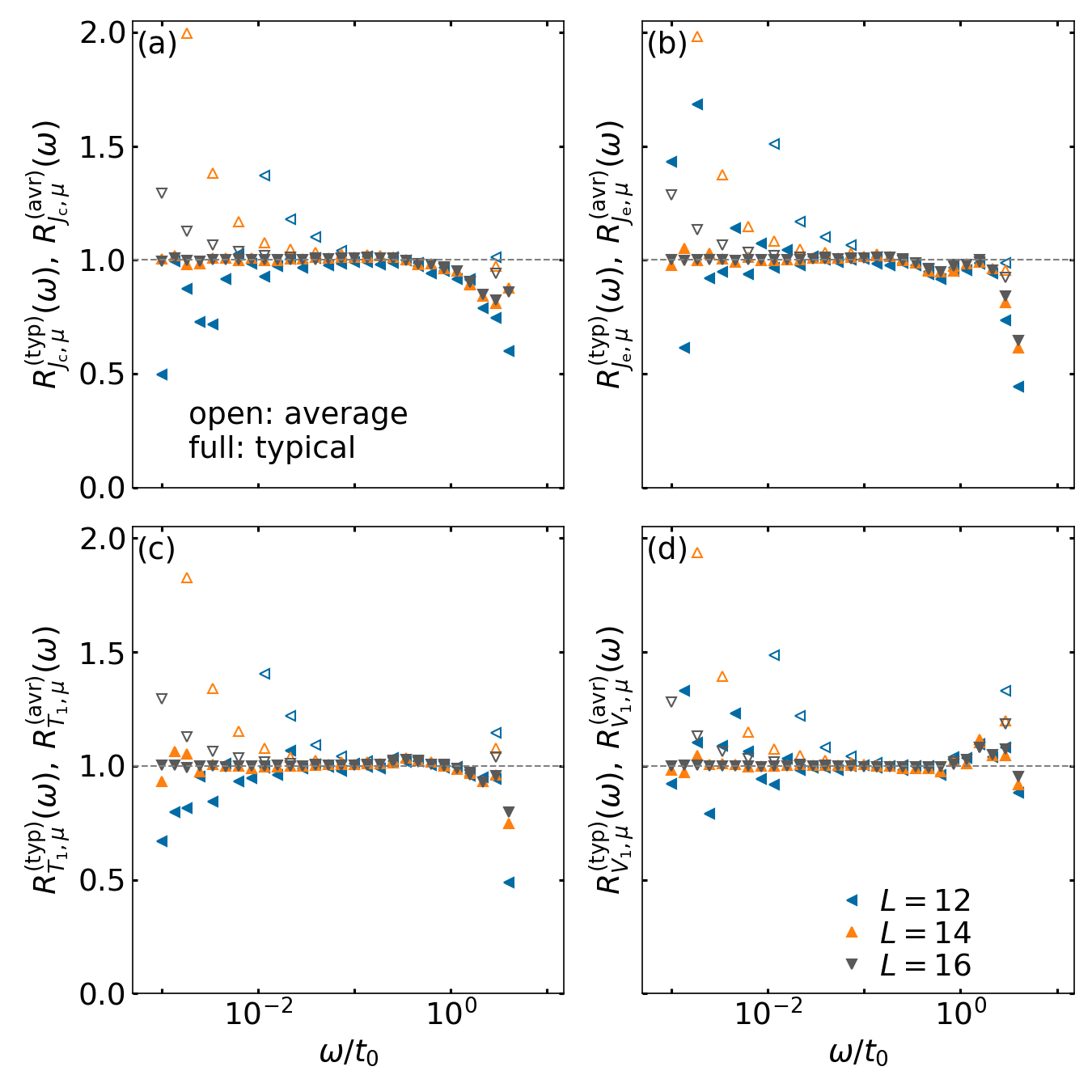}
\caption{
Numerical verification of the Kubo-Martin-Schwinger relation in Eq.~(\ref{def_kms}) for the means of ratios of the eigenstate spectral densities $S_O^{(\alpha)}(\omega)$ and $S_O^{(\alpha)}(-\omega)$.
Results are shown for the observables (a) $\hat J_{\rm c}$, (b) $\hat J_{\rm e}$, (c) $\hat T_1$ and (d) $\hat V_1$.
Horizontal lines are analytical predictions at $\beta = 0$, and symbols are numerical results for $L=12,14,16$.
Full symbols represent the typical values $R_{O,\mu}^{(\rm typ)}(\omega)$ from Eq.~(\ref{def_R_typical}) and open symbols represent the average values $R_{O,\mu}^{(\rm avr)}(\omega)$ from Eq.~(\ref{def_R_average}).
In Eqs.~(\ref{def_R_typical}) and~(\ref{def_R_average}), we choose $\mu$ such that the means include 10\% of all eigenstates w.r.t.~the center of the spectrum.
We discretize the $\omega/t_0$ axis by choosing 30 (15) points for $R_{O,\mu}^{(\rm typ)}(\omega) [R_{O,\mu}^{(\rm avr)}(\omega)]$ that are equidistant on a log scale from $10^{-3}$ to $10^1$.
}
\label{fig:kms_eigen}
\end{figure}

Results for both means are shown in Fig.~\ref{fig:kms_eigen} for the currents $\hat J_{\rm c}$ and $\hat J_{\rm e}$, and for the two spin observables $\hat T_1$ and $\hat V_1$.
The most striking feature of Fig.~\ref{fig:kms_eigen} is that at $L=16$ and $\omega/t_0 \lesssim 1$, the typical value of the ratio $R_{O,\mu}^{(\rm typ)}(\omega)$ fulfills the Kubo-Martin-Schwinger relation~(\ref{def_kms}) with high numerical accuracy.
These results demonstrate the validity of the FDT for the overwhelming majority of eigenstates in the bulk of the spectrum.

A more careful look reveals that, in fact, for $\omega/t_0 \ll 1$, both the typical and the average values exhibit a tendency to approach the expected asymptotic value by increasing $L$.
We find (not shown here) that the origin of the difference between the typical and the average values at $\omega/t_0 \to 0$ in finite systems stems from a small number of outliers that strongly differ from the means (the latter are calculated within a target interval on the $\omega$ axis, which is set by the width $\delta \omega$, as explained above).
The typical values are less sensitive to such fluctuations.
It would be interesting to test numerically whether {\it all} eigenstates in the bulk of the spectrum satisfy the FDT,  this goal, however,  is beyond the scope of the current work.

\begin{figure}[!]
\includegraphics[width=0.99\columnwidth]{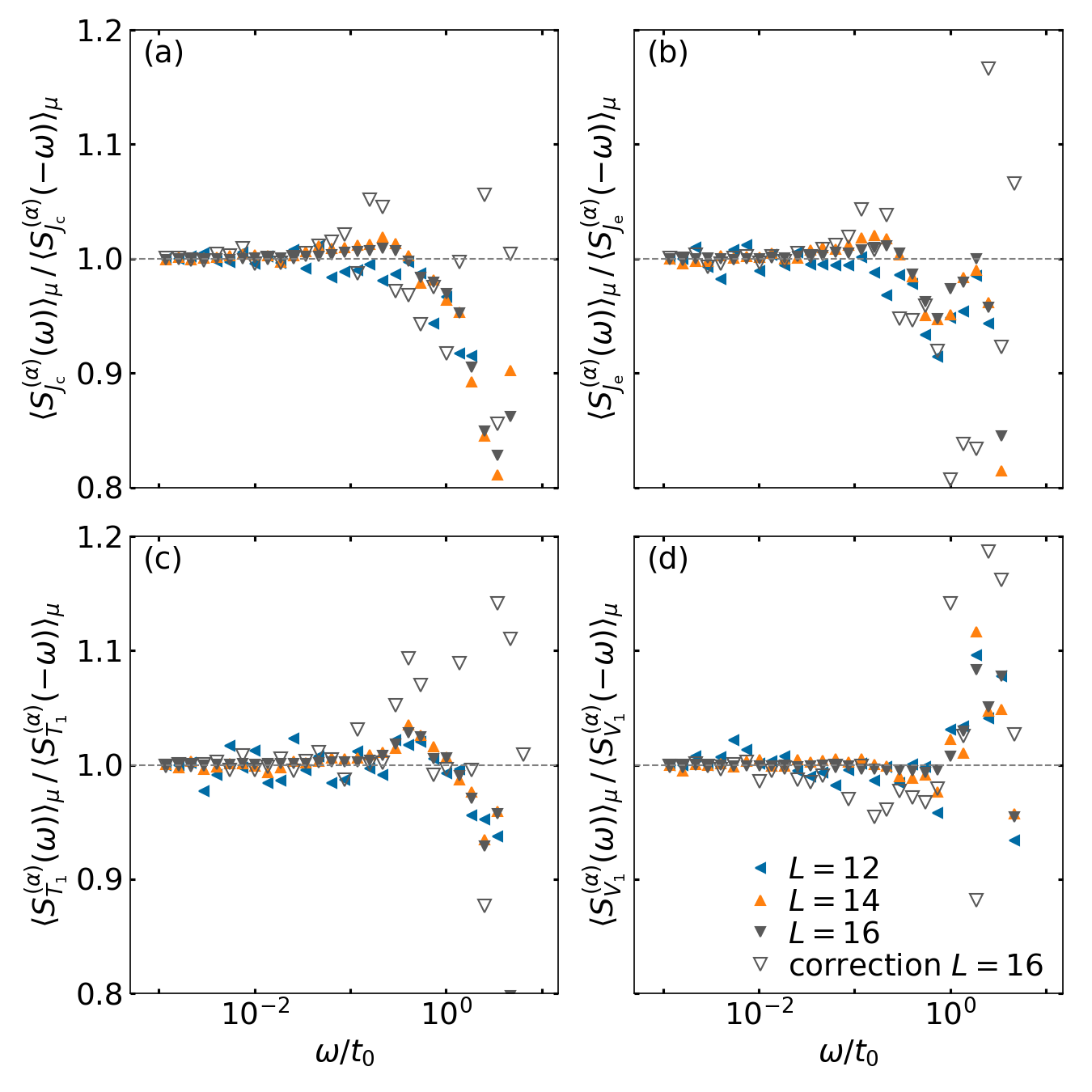}
\caption{
Ratios of averages of the eigenstate spectral densities $\langle S_O^{(\alpha)}(\omega) \rangle_\mu$ and $\langle S_O^{(\alpha)}(-\omega) \rangle_\mu$, defined in Eq.~(\ref{def_S_avr}).
Horizontal lines are results for the Kubo-Martin-Schwinger relation in Eq.~(\ref{def_kms}) at $\beta = 0$.
Filled symbols are numerical results for $L=12,14,16$.
We choose $\mu$ in Eq.~(\ref{def_S_avr}) such that the average includes 10\% of all eigenstates w.r.t.~the center of the spectrum.
We discretize the $\omega/t_0$ axis by choosing 30 points that are equidistant on a log scale from $10^{-3}$ to $10^1$.
Open symbols are numerical results for the finite-size correction $\exp\{ \omega \frac{\partial \ln|f_O(\bar E,\omega)|^2}{\partial \bar E} \}$, see Eq.~(\ref{def_ratio_2}), at $L=16$.
We calculate the discrete derivative over $\bar E$ by first calculating $|f_O(\bar E,\omega)|^2$ at $\bar E = 0$ and $\bar E = -\Delta$, where $\Delta/L = 0.0025$.
For each target $\bar E$, the results are averaged over all matrix elements for which the mean energy is in the interval $[\bar E -\Delta/2, \bar E + \Delta/2]$.
}
\label{fig:kms}
\end{figure}

A somehow different approach to the numerical verification of the FDT is to calculate the ratio of the averages of $S_O^{(\alpha)}(\omega)$ and $S_O^{(\alpha)}(-\omega)$, where the latter are defined as
\begin{equation} \label{def_S_avr}
 \langle S_O^{(\alpha)}(\pm\omega) \rangle_\mu = \frac{1}{\mu} \sum_{\alpha' = {\cal D}/2-\mu/2+1}^{{\cal D}/2+\mu/2} S_{O}^{(\alpha')}(\pm\omega) \,.
\end{equation}
The results are shown as filled symbols in Fig.~\ref{fig:kms}.
They exhibit certain similarities with the results for the means of the ratios in Fig.~\ref{fig:kms_eigen}.
In particular, all quantities under investigation match the Kubo-Martin-Schwinger prediction~(\ref{def_kms}) for $\omega/t_0 \lesssim 1$.
Note that the agreement between the average of the ratio and the ratio of averages, observed here for the spin-fermion model, is consistent with the recent results for  Heisenberg spin chains~\cite{noh_sagawa_20}.

To summarize our main numerical results, we observe a nearly perfect numerical agreement at small $\omega/t_0$ with predictions from the Kubo-Martin-Schwinger relation at $\beta =0$, which we express as
\begin{equation}
 R_{O,\mu}^{(\rm typ)}(\omega) \approx R_{O,\mu}^{(\rm avr)}(\omega) \approx \frac{\langle S_O(\omega) \rangle_\mu}{\langle S_O(-\omega) \rangle_\mu} \approx 1 \,, \;\;\; \omega/t_0 \ll 1 \,.
\end{equation}
These results show that at small $\omega/t_0$, the fluctuation-dissipation relation (without any finite-size corrections) can be observed with high accuracy for system sizes accessible with full exact diagonalization.
Another message from Figs.~\ref{fig:kms_eigen} and~\ref{fig:kms} is that the finite-size corrections become dominant at $\omega/t_0 \approx 1$ for all four observables studied in Fig.~\ref{fig:kms}.
However, the functional form of the corrections may strongly depend on the specific observable.

\subsection{Finite-size corrections}
\label{sec:fdt_fss}

It is interesting to analytically explore the leading finite-size corrections to the Kubo-Martin-Schwinger relation.
For the ratio of the eigenstate spectral densities, the corrections follow directly from Eq.~(\ref{def_S_alpha_expand}).
To the leading order, they give
\begin{align} \label{def_S_alpha_numerics}
 \frac{S_O^{(\alpha)}(\omega)} {S_O^{(\alpha)}(-\omega)} = & \frac{ |f_O(E_\alpha,\omega)|^2 + \frac{\omega}{2}\frac{\partial |f_O(E_\alpha,\omega)|^2}{\partial E_\alpha} }{|f_O(E_\alpha,\omega)|^2 - \frac{\omega}{2}\frac{\partial |f_O(E_\alpha,\omega)|^2}{\partial E_\alpha} } \, e^{\beta \omega} \nonumber \\
 \approx & \left(1 + \omega \frac{\partial \log( |f_O(E_\alpha,\omega)|^2)}{\partial E_\alpha} \right) e^{\beta \omega} \,.
\end{align}
However, the numerical evaluation of the latter expression, averaged over a window of eigenstates, is a difficult task.
Instead, it was argued that the finite-size corrections are more easily treated for the ratio of averages of the eigenstate spectral densities~\cite{noh_sagawa_20}.
For the latter, the leading term can be expressed as a function of $\bar E$, where $\bar E$ is the mean energy of eigenstates in the microcanonical window with an energy variance $\sigma^2$.
The average eigenstate spectral density is then~\cite{noh_sagawa_20}
\begin{align}
 \langle S_O^{(\alpha)}(\pm\omega)\rangle_\mu = & \; 2\pi |f_O(\bar E,\omega)|^2 \, e^{\pm \frac{\beta \omega}{2} + \frac{3\omega^2}{8}\frac{\partial \beta}{\partial \bar E}} \nonumber \\
 & \times e^{ \pm \frac{\omega}{2} \frac{\partial \ln|f_O(\bar E,\omega)|^2}{\partial \bar E} + {\cal O}(\sigma^2/\bar E^2) + ... }
 \label{def_ratio_1}
\end{align}
and hence the ratio of averages is
\begin{equation}
 \frac{\langle S_O^{(\alpha)}(\omega)\rangle_\mu}{\langle S_O^{(\alpha)}(-\omega)\rangle_\mu} = e^{ \beta \omega + \omega \frac{\partial \ln|f_O(\bar E,\omega)|^2}{\partial \bar E} + {\cal O}(\sigma^2/\bar E^2) + ... } \, .
 \label{def_ratio_2}
\end{equation}
Note that there are two sources of finite-size corrections in Eqs.~(\ref{def_ratio_1}) and~(\ref{def_ratio_2}).
The first stems from the expansion of quantities around the mean energy $\bar E$.
An example of such a contribution in Eq.~(\ref{def_ratio_2}) is $\omega \partial \ln|f_O(\bar E,\omega)|^2/\partial \bar E$, which, for large systems, scales as $\propto 1/\bar E \approx 1/L$.
The higher-order terms not given in Eq.~(\ref{def_ratio_2}) are at most of the order $\propto 1/\bar E^2 \approx 1/L^2$.
The second stems from the width of the microcanonical window in which the results are averaged, and it is governed by the microcanonical variance $\sigma^2$.
The leading term of this contribution is denoted as ${\cal O}(\sigma^2/\bar E^2)$ in Eq.~(\ref{def_ratio_2}), and scales as $\propto L/L^2 = 1/L$.
It is therefore of the same order as the term $\omega \partial \ln|f_O(\bar E,\omega)|^2/\partial \bar E$.
However, by choosing narrow microcanonical windows, we numerically verified (not shown here) that the impact of the finite width of the microcanonical window is negligible, and hence not explicitly considered in Eqs.~(\ref{def_ratio_1}) and~(\ref{def_ratio_2}).

Open symbols in Fig.~\ref{fig:kms} show the correction to the asymptotic value evaluated as $\exp\{ \omega \frac{\partial \ln|f_O(\bar E,\omega)|^2}{\partial \bar E} \}$, see Eq.~(\ref{def_ratio_2}).
The results show that the correction (open symbols) roughly agrees with the numerical finite-size  results (full symbols).
However, the quantitative agreement is not very accurate.
This suggests that for large $\omega/t_0$, other contributions neglected in the derivation of Eq.~(\ref{def_ratio_2}) should also be relevant.

\section{Autocorrelation functions}
\label{sec:autocorrelation}

In this section, we explore what can be learned about the properties of the observable matrix elements $|O_{\alpha\beta}|^2$ from the autocorrelation functions $C_O^{(\alpha)}(t)$, defined in Eq.~(\ref{def_C_alpha}), and the corresponding spectral densities.

\subsection{Accessing spectral properties from the integrated autocorrelation functions}

Here, we study time integrals of the autocorrelation functions, with the focus on the currents.
The emphasis is given to the properties at long times.
At short times (not shown here), we find that the charge-current autocorrelation function $C_{J_{\rm c}}^{(\alpha)}(t)$ typically decays as a Lorentzian function, which is consistent with the exponential decay of the corresponding offdiagonal matrix elements reported in Fig.~\ref{fig:offdiag_current}(a).

The time integral of the autocorrelation function $C_O^{(\alpha)}(t)$ from Eq.~(\ref{def_C_alpha_matele}) is
\begin{align}
 D_O^{(\alpha)} (t) & = \int_{-t}^{t} C_O^{(\alpha)} (t') dt' = 4 \sum_{\beta \neq \alpha} |O_{\alpha\beta}|^2 \frac{\sin[(E_\beta - E_\alpha)t]}{E_\beta - E_\alpha} \nonumber \\
 & = 4\pi \int_{-\infty}^\infty d\omega \rho(E_\alpha+\omega) |O_{E_\alpha,E_\alpha+\omega}|^2 \;  \frac{\sin(\omega t)}{\pi\omega} \,,
 \label{def_D_alpha}
\end{align}
where in the second row, the sum over eigenstates was replaced by the integral.
If the time is large enough, one can replace $\sin(\omega t)/(\pi \omega) \to \delta(\omega)$.
This is indeed a reasonable approximation for times $t > 1/\Omega^*$, where $\Omega^*$ is chosen such that the density of states $\rho(E_\alpha+\omega)$ and the matrix elements $|O_{E_\alpha,E_\alpha+\omega}|^2$ are independent of $\omega$ at $\omega < \Omega^*$.
In this time regime, we get
\begin{equation} \label{agreement_D_S}
 D_O^{(\alpha)} \left(t > \frac{1}{\Omega^*}\right) = 4\pi \overline{|O_{\alpha\alpha'}|^2} \rho(E_\alpha) = S_{O,+}^{(\alpha)}(\omega \approx 0) \,.
\end{equation}
Hence, the long-time limit of the integrated autocorrelation function agrees with the $\omega\to0$ result of the spectral density $S_{O,+}^{(\alpha)}(\omega)$ introduced in Eqs.~(\ref{def_S_alpha}) and~(\ref{def_S_alpha_plus}).

\begin{figure}[!]
\includegraphics[width=1.0\columnwidth]{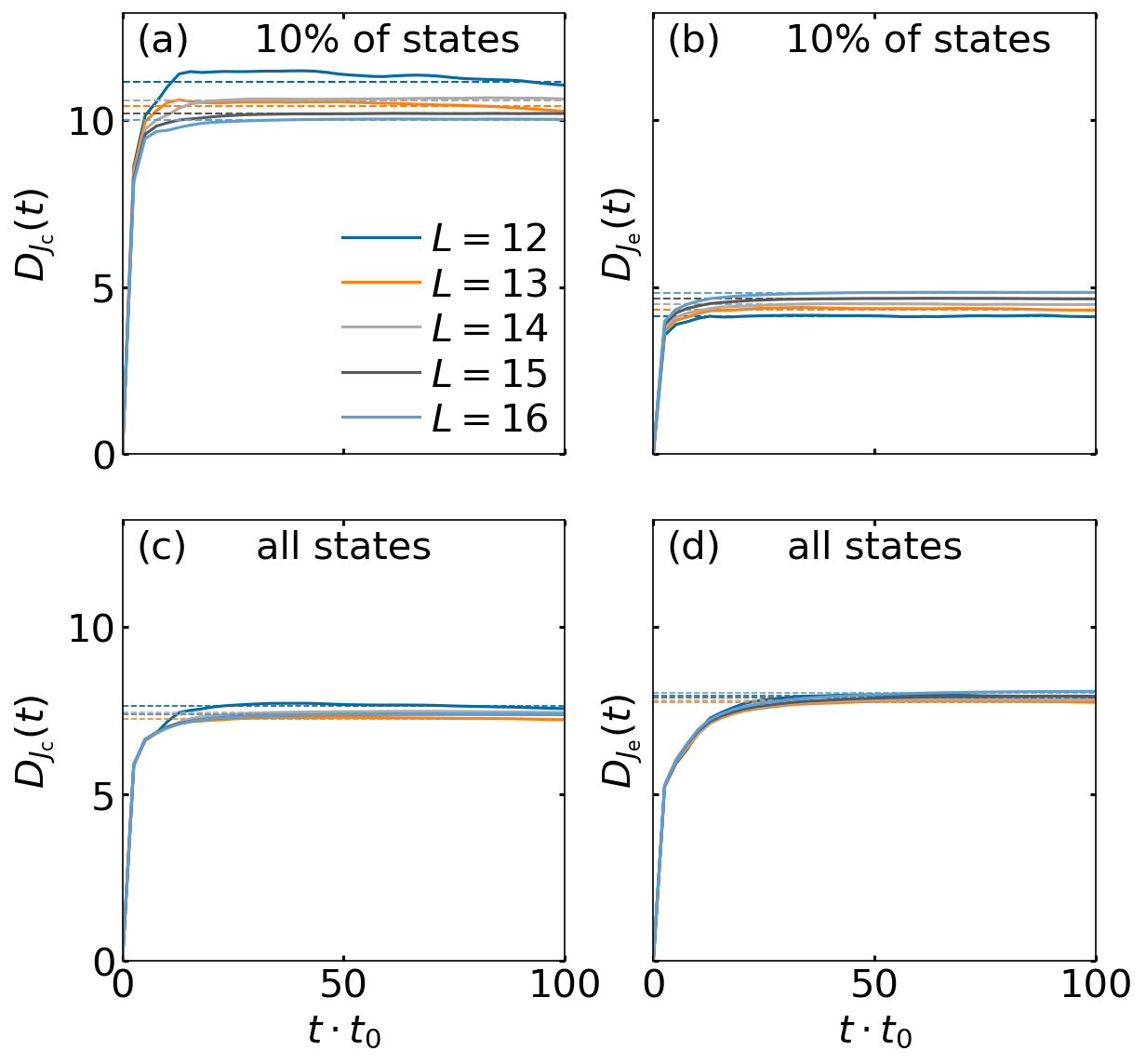}
\caption{
Time integration of the autocorrelation functions.
Solid lines: Integrated autocorrelation functions $D_O^{(\alpha)}(t)$ [see the first row in Eq.~(\ref{def_D_alpha})], averaged over [(a),(b)] $10\%$ of eigenstates $|\alpha\rangle$ in the middle of the spectrum, and [(c),(d)] over all eigenstates.
Observables are the charge current $\hat J_{\rm c}$ [(a),(c)] and the energy current $\hat J_{\rm e}$ [(b),(d)].
The horizontal dashed lines represent the $\omega/t_0\to 0$ limits of the spectral densities, evaluated for the corresponding matrix elements of the operators and the density of states as given by the r.h.s.~of Eq.~(\ref{agreement_D_S}).
}
\label{fig:current_integrated}
\end{figure}

\begin{figure*}[!]
\includegraphics[width=2.0\columnwidth]{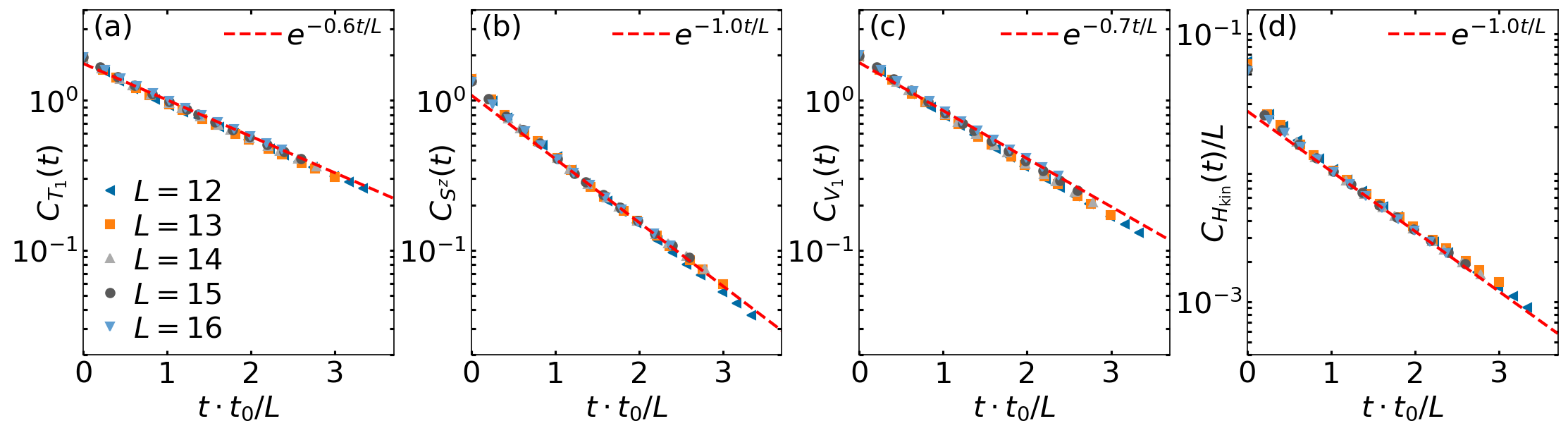}
\caption{
Time evolution [(a)-(c)] of the autocorrelation functions $C_O(t) = \langle C_O^{(\alpha)}(t) \rangle_\mu$ for the observables $\hat T_1$, $\hat S^z$ and $\hat V_1$, respectively,
and [(d)] of the scaled autocorrelation function $C_O(t)/L$ for the kinetic energy $\hat H_{\rm kin}$.
The latter scaling is consistent with the additional division by $L$ of the offdiagonal matrix elements of $\hat H_{\rm kin}$ in Fig.~\ref{fig:offdiag_linear}(d).
We choose $\mu$ such that $C_O^{(\alpha)}(t)$ [see Eq.~(\ref{def_C_alpha_matele})] is averaged over all eigenstates $|\alpha\rangle$.
Results are shown for different system sizes $L$ (see legend) versus the scaled time $t t_0/L$.
The dashed line is a fit to the $L=16$ results for $tt_0/L > 1$.
}
\label{fig:offdiag_exp}
\end{figure*}

In general, the time $t^* = 1/\Omega^*$ needed for the integrated autocorrelation function to match the $\omega\to0$ limit of the spectral density may be very large provided that usually, $1/\Omega^* \propto L^\nu$, with $\nu > 0$.
However, based on our analysis carried out in Sec.~\ref{sec:structure} for the currents, we expect that $\Omega^*$ does not scale with $L$ for system sizes under investigation, and hence  Eq.~(\ref{agreement_D_S}) should be valid already at moderately short times.
We verify this explicitly for both charge and energy currents: the solid lines in Fig.~\ref{fig:current_integrated} represent the numerical evaluation of the first line in Eq.~(\ref{def_D_alpha}) in a microcanonical window, and the results indeed approach a time-independent value after a relatively short time.
A comparison of the averages in different microcanonical windows around the center of the spectrum [10\% of states in Figs.~\ref{fig:current_integrated}(a)-\ref{fig:current_integrated}(b) versus all states in Figs.~\ref{fig:current_integrated}(c)-\ref{fig:current_integrated}(d)] suggests that for the averages in very large windows, the system-size dependence becomes negligible.

In Fig.~\ref{fig:current_integrated}, we also numerically verify Eq.~(\ref{agreement_D_S}).
In particular, we predict the long-time values of the time-integrated autocorrelation functions (solid lines) by calculating the mean of the squared offdiagonal matrix elements in the limit $\omega/t_0 \to 0$ (horizontal dashed lines).
The latter are calculated using Eq.~(\ref{def_S1_alpha}) and we average all quantities in microcanonical windows.
The agreement between the two results is excellent.

The results shown in Fig.~\ref{fig:current_integrated} demonstrate the possibility of extracting the low-$\omega$ properties of the offdiagonal matrix elements from numerically calculating the autocorrelation functions.
This could be exploited by numerical methods beyond exact diagonalization (e.g., matrix-product state methods~\cite{karrasch_bardarson_12, paeckel_koehler_19}, dynamical quantum typicality~\cite{stenigeweg_gemmer_14, stenigeweg_heidrichmeisner_14}, and the numerical linked cluster expansion~\cite{richter_steinigeweg_19, mallayya_rigol_18, guardadosanchez_brown_18, white_sundar_17}), which allow for an efficient time evolution of pure quantum states that correspond to a target energy window.
For the currents, note that a proper statistical ensemble average of $D_O^{(\alpha)}(t>1/\Omega^*)$ is related to the dc conductivity and the diffusion constant, as exploited in a series of studies of transport in spin models~\cite{steinigeweg_gemmer_09, steinigeweg_brenig_11, bertini_heidrichmeisner_21}.
Specifically, at infinite temperature ($\beta \to 0$ limit), one can express the regular part of the optical conductivity as~\cite{mahan_00, jaklic_prelovsek_00}
\begin{equation}
 \sigma_{\rm reg}^{\infty} (\omega) = \beta \; \mathbb{R} \left\{ \int_0^\infty dt e^{i\omega t} \frac{1}{\cal D} \sum_\alpha \langle \alpha | \hat J_{\rm c}(t) \hat J_{\rm c}(0) | \alpha\rangle \right\} \,,
\end{equation}
which gives
\begin{equation}
 \beta^{-1} \sigma_{\rm reg}^\infty (\omega) = \frac{\pi}{\cal D} \sum_\alpha \overline{|(J_{\rm c})_{E_\alpha, E_\alpha+\omega}|^2} \, \rho(E_\alpha + \omega) \,.
\end{equation}
Hence, the statistical ensemble average of $S_{J_{\rm c}}^{(\alpha)}(\omega \approx 0)$ in the $\beta \to 0$ limit yields
\begin{equation}
 \langle S_{J_{\rm c}}^{(\alpha)} (\omega \approx 0) \rangle_{\beta \to 0} = 4 \lim_{\omega\to 0} \beta^{-1} \sigma_{\rm reg}^\infty (\omega) \,.
\end{equation}
The latter quantity is shown as horizontal dashed lines in Fig.~\ref{fig:current_integrated}(c).

\subsection{Autocorrelation functions of observables with Drude-like structure}
\label{sec:current_auto}

Finally, we focus on the class of observables for which the scaled offdiagonal matrix elements exhibit a Drude-like decay in a broad energy window, described by Eq.~(\ref{def_lorentzian}).
A Fourier transform of the latter function, which is proportional to the autocorrelation function, then decays exponentially with time,
\begin{equation} \label{C_alpha_exp}
 C_O^{(\alpha)} (t) \propto e^{-\mu_O t} \,.
\end{equation}
Such a time dependence is indeed observed in microcanonical averages of the autocorrelation functions for the observables $\hat T_1$, $\hat S^z$, $\hat V_1$ and $\hat H_{\rm kin}$, as shown in Fig.~\ref{fig:offdiag_exp}.

Apart from the mere observation of the exponential decay of the autocorrelation functions, it is also interesting to study the $L$ dependence of the rate $\mu_O$.
The Fourier transform of the function in Eq.~(\ref{def_lorentzian}) predicts that the rate is inversely proportional to $L$.
In Fig.~\ref{fig:offdiag_exp}, we plot the autocorrelation functions as functions of the scaled time $t t_0/L$, which yields an excellent data collapse for different system sizes. 
These results are therefore consistent with the observations made by solely studying the offdiagonal matrix elements.

The $L$ dependence of the rate suggests that the autocorrelation functions decay to zero only after extensively long times, $t \gtrsim L/t_0$.
The linear dependence on $L$ may suggest fingerprints of ballistic transport and proximity to an integrable point (which is realized if the spin-fermion coupling vanishes).
However, some care is needed in this interpretation since the rate is calculated in the limit of a vanishing fermion occupation density, $n_{\rm f} \propto 1/L$.
Hence further studies are needed to shed more light onto the signatures of ballistic versus diffusive propagation in isolated quantum systems with itinerant impurities.

\section{Conclusions} \label{sec:conclusion}

ETH is a well-established framework whose applicability has been demonstrated for various quantum chaotic models.
Several intriguing aspects of the ETH ansatz have recently been explored, to list just a few of them:
(i) its validity close to integrable points~\cite{jansen_stolpp_19, brenes_leblond_20, brenes_goold_20, santos_perezbernal_20} or in few-body systems~\cite{zisling_santos_21, fogarty_garciamarch_21};
(ii) existence of a well-defined variance of the offdiagonal matrix elements of observables in the interacting integrable Heisenberg model, which is a function of the average energy and the energy difference~\cite{leblond_mallayya_19, mallayya_rigol_19}, in analogy to the structure function $|f_O(\bar E, \omega)|^2$ in the ETH ansatz~(\ref{def_eth_ansatz});
(iii) characterization of different classes of observables, e.g., their relation to conserved quantities~\cite{mierzejewski_vidmar_20} and the impact of observables that break the symmetry of the Hamiltonian~\cite{leblond_rigol_20},
(iv) validity of the fluctuation-dissipation relation for eigenstates in finite systems~\cite{dalessio_kafri_16, noh_sagawa_20}, and
(v) the structure of four-point functions within the ETH ansatz~\cite{foini_kurchan_19, chan_deluca_19, murthy_srednicki_19}.

The main results of our work are aligned with several of those questions.
First, a single itinerant fermion, coupled to an integrable spin-1/2 XX chain, restores ergodicity and gives rise to the validity of the ETH ansatz as was shown in \cite{jansen_stolpp_19} and studied in more detail here.
In fact, studying the fluctuations of the matrix elements of translationally invariant observables in this model, we showed that the ETH is fulfilled to remarkable numerical accuracy, which can serve as a benchmark for future studies.
Our study highlights that the scaling analysis of several ETH indicators is most conveniently carried out for structureless operators, i.e., the operators for which $O(E) \approx 0$ in the ETH ansatz~(\ref{def_eth_ansatz}).
Those operators have no overlap with the Hamilton operator, in the sense defined in Ref.~\cite{mierzejewski_vidmar_20}.

We then focused on the observable-dependent structure of the offdiagonal matrix elements, and we complemented the analysis by studying the autocorrelation functions and their time integrals.
Most importantly, we showed that several key features of the offdiagonal matrix elements of different classes of observables can be detected from the autocorrelation functions, which opens the door for future studies of the matrix elements of observables using numerical techniques beyond full exact diagonalization.
A special attention was devoted to two classes of observables, which have been rarely studied before:
(i) the charge and energy current operators, whose system-size dependence of the scaled offdiagonal matrix elements $\overline{|O_{\alpha\beta}|^2} {\cal D}$ for eigenstates in the bulk of the spectrum strongly differs from all other observables under investigation, and
(ii) a class of observables with Drude-like $\omega$-dependence.

Finally, we numerically explored the validity of the fluctuation-dissipation relation for the eigenstates in the bulk of the spectrum. 
We identified the regime of energies $\omega$ that are lower than the typical values of model parameters, for which the fluctuation-dissipation relation is fulfilled with high accuracy in systems that can be studied with full exact diagonalization.
At higher energies, however, the relation is governed by subleading terms that in finite systems do not necessarily exhibit a convergence to the result in the thermodynamic limit.

\section*{Acknowledgement} \label{sec:ack}

We acknowledge useful discussions with S. Kehrein, M. Mierzejewski, A. Polkovnikov, P. Prelov\v{s}ek and M. Rigol. 
This work was supported by the Deutsche Forschungsgemeinschaft (DFG, German Research Foundation)   -- 207383564; 217133147;  via Research Unit FOR 1807 and CRC 1073 (project B09), respectively.
L.V. acknowledges support from the Slovenian Research Agency (ARRS), Research core fundings Grants No.~P1-0044 and No.~J1-1696.




\bibliographystyle{biblev1}
\bibliography{references}

\end{document}